\pdfoutput=1
\documentclass[11pt,twoside,a4paper,cmspaper,final,collab]{cms-tdr}
\def\svnVersion{d995e27}\def\svnDate{2024/08/26}\def\cmsCernNoTag{CERN-EP-2024-127}\def\cmsCernDate{\today}\def\cmsMessage{Published in Reports on Progress in Physics as \href{http://dx.doi.org/10.1088/1361-6633/ad6fcb}{\doi{10.1088/1361-6633/ad6fcb}.}}
\begin{document}\cmsNoteHeader{SMP-23-005}

\newcommand{\mvis}{\ensuremath{m_\text{vis}}\xspace}
\newcommand{\at}{\ensuremath{a_{\PGt}}\xspace}
\newcommand{\dt}{\ensuremath{d_{\PGt}}\xspace}
\newcommand{\mtt}{\ensuremath{m_{\PGt\PGt}}\xspace}
\newcommand{\mmumu}{\ensuremath{m_{\PGm\PGm}}\xspace}
\newcommand{\mll}{\ensuremath{m_{\Pell\Pell}}\xspace}
\newcommand{\mZ}{\ensuremath{m_{\PZ}}\xspace}
\newcommand{\ntracks}{\ensuremath{N_\text{tracks}}\xspace}
\newcommand{\ntracksHS}{\ensuremath{N_\text{tracks}^\text{HS}}\xspace}
\newcommand{\ntracksPU}{\ensuremath{N_\text{tracks}^\text{PU}}\xspace}
\newcommand{\ggtotautau}{{\ensuremath{\PGg\PGg\to\PGt\PGt}}\xspace}
\newcommand{\ggtomumu}{{\ensuremath{\PGg\PGg\to\PGm\PGm}}\xspace}
\newcommand{\ggtoee}{{\ensuremath{\PGg\PGg\to\Pe\Pe}}\xspace}
\newcommand{\ggtoww}{{\ensuremath{\PGg\PGg\to\PW\PW}}\xspace}
\newcommand{\Wjets}{\ensuremath{\PW+\text{jets}}\xspace}
\newcommand{\Zjets}{\ensuremath{\PZ/\PGg^{*}+\text{jets}}\xspace}
\newcommand{\mumu}{\ensuremath{\PGm\PGm}\xspace}
\newcommand{\emu}{\ensuremath{\Pe\PGm}\xspace}
\newcommand{\etau}{\ensuremath{\Pe\tauh}\xspace}
\newcommand{\mutau}{\ensuremath{\PGm\tauh}\xspace}
\newcommand{\tautau}{\ensuremath{\tauh\tauh}\xspace}
\newcommand{\DMtauh}{\ensuremath{\text{DM}^{\tauh}}\xspace}
\newcommand{\pttauh}{\ensuremath{\pt^{\tauh}}\xspace}
\newcommand{\MFcorr}{\ensuremath{\text{MF}^\text{corr}}\xspace}
\newcommand{\MFtot}{\ensuremath{\text{MF}^\text{tot}}\xspace}
\newcommand{\NW}{\ensuremath{N^{\PW}}\xspace}
\newcommand{\NQCD}{\ensuremath{N^\text{QCD}}\xspace}
\newcommand{\xQCD}{\ensuremath{x^\text{QCD}}\xspace}
\newcommand{\zBS}{\ensuremath{z_\text{BS}}\xspace}
\newcommand{\sBS}{\ensuremath{\sigma_\text{BS}}\xspace}
\newcommand{\CTB}{\ensuremath{C_{\PGt B}}\xspace}
\newcommand{\CTW}{\ensuremath{C_{\PGt W}}\xspace}
\newcommand{\CTg}{\ensuremath{C_{\PGt\PGg}}\xspace}
\newcommand{\cmns}{\ensuremath{\text{cm}}\xspace} 
\newcommand{\ecm}{\,\ensuremath{e}\cm\xspace}
\renewcommand{\Re}[1]{\ensuremath{\operatorname{Re}\big[#1\big]}}
\renewcommand{\Im}[1]{\ensuremath{\operatorname{Im}\big[#1\big]}}

\newlength\cmsFigWidth\setlength\cmsFigWidth{0.45\textwidth}
\newlength\cmsTabSkip\setlength{\cmsTabSkip}{1ex}
\ifthenelse{\boolean{cms@external}}{\providecommand{\cmsLeft}{upper\xspace}}{\providecommand{\cmsLeft}{left\xspace}}
\ifthenelse{\boolean{cms@external}}{\providecommand{\cmsRight}{lower\xspace}}{\providecommand{\cmsRight}{right\xspace}}
\ifthenelse{\boolean{cms@external}}{\providecommand{\cmsUpperLeft}{upper\xspace}}{\providecommand{\cmsUpperLeft}{upper left\xspace}}
\ifthenelse{\boolean{cms@external}}{\providecommand{\cmsUpperRight}{middle\xspace}}{\providecommand{\cmsUpperRight}{upper right\xspace}}

\cmsNoteHeader{SMP-23-005}
\title{Observation of \texorpdfstring{$\ggtotautau$}{gamma gamma to tau tau} in proton-proton collisions and limits on the anomalous electromagnetic moments of the \texorpdfstring{\PGt}{tau} lepton}
\titlerunning{Observation of $\ggtotautau$ in proton-proton collisions}

\date{\today}

\abstract{
The production of a pair of $\PGt$ leptons via photon-photon fusion, $\ggtotautau$, is observed for the first time in proton-proton collisions, with a significance of 5.3 standard deviations. This observation is based on a data set recorded with the CMS detector at the LHC at a center-of-mass energy of 13\TeV and corresponding to an integrated luminosity of 138\fbinv. Events with a pair of $\PGt$ leptons produced via photon-photon fusion are selected by requiring them to be back-to-back in the azimuthal direction and to have a minimum number of charged hadrons associated with their production vertex. The $\PGt$ leptons are reconstructed in their leptonic and hadronic decay modes. The measured fiducial cross section of $\ggtotautau$ is $\sigma^\text{fid}_\text{obs}=12.4 ^{+3.8}_{-3.1}\unit{fb}$. Constraints are set on the contributions to the anomalous magnetic moment (\at) and electric dipole moments (\dt) of the $\PGt$ lepton originating from potential effects of new physics on the $\PGg\PGt\PGt$ vertex: $\at= 0.0009_{-0.0031}^{+0.0032}$ and $\abs{\dt} < 2.9\times 10^{-17}\ecm$ (95\% confidence level), consistent with the standard model.
}

\hypersetup{
pdfauthor={CMS Collaboration},
pdftitle={Observation of the gamma gamma to tau tau production in proton-proton collisions at sqrt(s)=13 TeV and limits on the anomalous electromagnetic moments of the tau lepton},
pdfsubject={CMS},
pdfkeywords={CMS, exclusive production, taus}}

\maketitle

\section{Introduction}

The production of lepton pairs via photon-photon fusion in proton-proton ($\Pp\Pp$) collisions draws significant interest since it is a pure quantum electrodynamics~(QED) process at leading order, for which the theoretical cross section 
can be calculated with uncertainties below 1\%~\cite{Budnev:1973tz}. If both protons remain intact after their electromagnetic interaction, the final state has a particularly clean topology characterized by two leptons emitted back-to-back in azimuth and without any hadronic activity surrounding it. Although the photon-fusion production of dileptons in hadronic collisions has been measured precisely for electrons and muons~\cite{STAR:2004bzo,CDF:2006apx,CDF:2009xey,CDF:2009xwv,ATLAS:2020epq,CMS:2012cve,CMS:2011vma}, a similar experimental precision is lacking for $\PGt$ leptons~\cite{ATLAS:2022ryk,CMS:2022arf}. The photon-fusion channel has long been proposed to probe the $\PGg-\PGt$ coupling at hadron colliders~\cite{delAguila:1991rm,Atag:2010ja,Dyndal:2020yen,PhysRevD.102.113008}, because the short $\PGt$ lepton lifetime makes other measurements difficult. 

The most general form of photon-lepton coupling ($ie\Gamma^{\mu}$) satisfying Lorentz invariance is:
\begin{equation}\label{eq:gamma_mu} 
  \Gamma^{\mu} = \gamma^\mu F_1\left(q^2\right) + \frac{\sigma^{\mu\nu}q_\nu}{2m}\left[iF_2\left(q^2\right) + F_3\left(q^2\right)\gamma_5\right],
\end{equation}
where $\sigma^{\mu\nu}=i[\gamma_\mu,\gamma_\nu]/2$ is the spin tensor proportional to the commutator of 
the gamma matrices, $q$ is the momentum transfer between incoming and outgoing leptons of mass $m$, and $F_i\left(q^2\right)$ ($i=1,2,3$) are form factors that contain the complete information about the electric and magnetic couplings of the lepton. 
Beyond serving as a sensitive test of QED interactions, the photon-photon production process also offers insights into the anomalous electromagnetic moments of the lepton 
($\Pell=\Pe,\Pgm,\PGt$), 
$F_2\left(0\right)=a_{\Pell}\equiv(g_{\Pell}-2)/2$ and $F_3\left(0\right)=-2 m d_{\Pell} / e$, and is a powerful probe of physics beyond the standard model (BSM). 
In this expression, the gyromagnetic 
ratio $g_{\Pell}$ is a constant term that relates the magnetic moment of the lepton to its spin, and $d_{\Pell}$ is the lepton anomalous electric dipole moment. 
The one-loop contribution to $a_{\Pell}$ is equal to the so-called "Schwinger term", which is common to all leptons~\cite{PhysRev.73.416}:
\begin{equation}
  a_{\Pell} = \frac{\alpha}{2\pi} \simeq 0.00116,
\end{equation}
where $\alpha$ is the fine-structure constant. 

The anomalous magnetic moment of the electron has been 
measured to agree with the standard model (SM) with an impressive precision of twelve significant digits~\cite{Parker_2018,Fan:2022eto}. 
Measurements of $a_\mu$ by the Muon g-2 Collaboration reach a precision of ten significant digits and appear in conflict with certain theoretical predictions~\cite{Muong-2:2021ojo,Muong-2:2023cdq,Aoyama:2020ynm}. 
If BSM effects scale with the squared lepton mass, deviations of $\at$ from the SM predictions would be more than two orders of 
magnitude larger than they are for $a_\mu$. The photon-fusion production of a pair of $\PGt$ leptons probes the $\PGt\PGt\PGg$ vertex 
and can be used to measure $\at$, which is predicted to be $\at = 1.17721\pm0.00005\times10^{-3}$ in the SM~\cite{Eidelman:2007sb}.
This same process can also be studied to constrain $\dt$, which, since there is no appreciable violation of charge-conjugation and parity symmetries in the charged lepton sector of the SM, is predicted to have a nonzero but tiny value~\cite{Yamaguchi:2020eub}.

\begin{figure}[ht]
  \centering
  \includegraphics[width=0.3\textwidth]{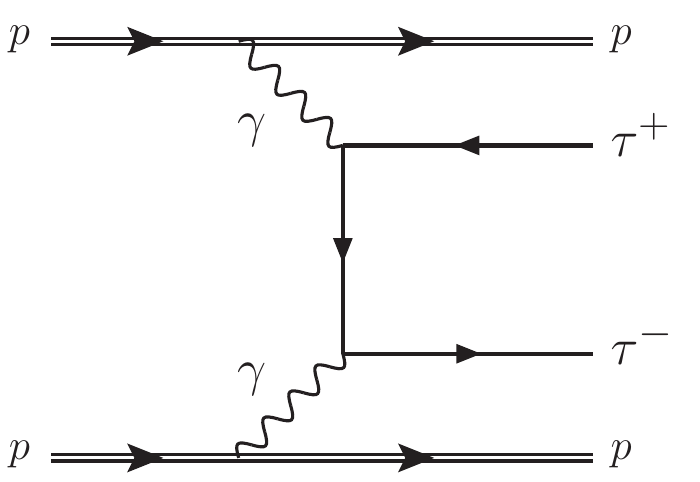}
  \includegraphics[width=0.3\textwidth]{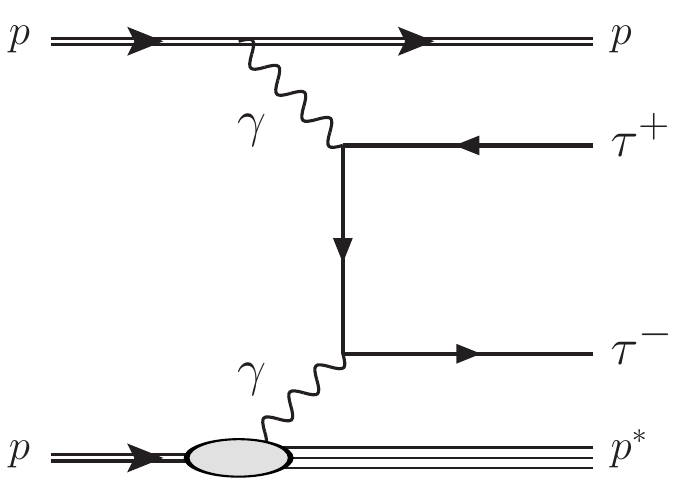}
  \includegraphics[width=0.3\textwidth]{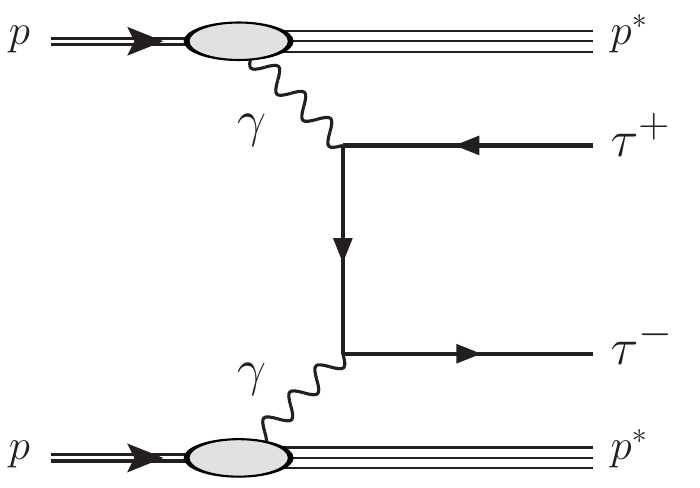}
  \caption{Feynman diagrams for the production of $\PGt$ lepton pairs by photon-photon fusion. The exclusive (\cmsLeft), single proton dissociation (middle), and double proton dissociation (\cmsRight) topologies are shown.}\label{fig:feynman}
\end{figure}

The signal in this analysis is $\Pp\Pp \xrightarrow{\PGg\PGg} \Pp^{(*)} \PGt^+ \PGt^- \Pp^{(*)}$, where $\Pp^{(*)}$
denotes a final-state proton that stays intact or fragments after the photon emission.  When both protons radiate coherently a photon and remain intact after their interaction, the process is called exclusive or elastic. Since the protons do not dissociate, 
these events are characterized by the absence of any hadron produced, other than those from the $\PGt$ lepton decays.
The photons emitted in the elastic processes are almost on-shell, limiting their virtuality to small values ($q^2\to 0$)~\cite{Budnev:1973tz}, and the 
produced $\PGt$ leptons are back-to-back in the azimuthal direction. 
When one or both incoming protons fragment as a result of photon emission, the process is called either single- or
double-dissociative. Single- and
double-dissociative events can also give rise to an elastic-like signature when the proton remnants are not
reconstructed because they fall outside of the detector acceptance.
Figure~\ref{fig:feynman} shows the three diagrams considered here, contributing to the $\ggtotautau$ process in $\Pp\Pp$ collisions. 

The $\ggtotautau$ process has been observed by the ATLAS and CMS Collaborations in ultraperipheral
collisions of lead ions~\cite{ATLAS:2022ryk,CMS:2022arf}, and used to set constraints on $\at$.
The observation was driven by signal events with a di-$\PGt$ invariant mass, $\mtt$, typically below 20\GeV.
This paper describes the first measurement of the photon-fusion production of a pair of $\PGt$ leptons in $\Pp\Pp$ collisions.
Because of the trigger requirements, experimentally accessible signal events have $\mtt>50\GeV$. The $\PGt\PGt$ production in this phase space has a much 
lower cross section, but the integrated luminosity is much higher than the one achievable with heavy ion collisions at the LHC. Additionally, BSM effects from new heavy particles~\cite{Crivellin:2021spu} are enhanced at
high mass, and the phase space accessible in $\Pp\Pp$ collisions is more sensitive to BSM scenarios with non-SM $\at$ and $\dt$ 
values than that accessible in ultraperipheral
collisions of lead ions. 
Outside of LHC experiments, constraints on $\at$ were previously set by the DELPHI, OPAL, and L3 experiments~\cite{DELPHI:2003nah,OPAL:1998dsa,L3:1998169}. 
Constraints on $\dt$ were determined by the Belle, L3, OPAL, and ARGUS Collaborations~\cite{Belle:2021ybo,ARGUS:2000riz,OPAL:1998dsa,L3:1998169}. 

This paper is based on $\Pp\Pp$ collision data at $\sqrt{s}=13\TeV$ collected with the CMS detector in 2016--2018, corresponding to an integrated luminosity of 138\fbinv.
To select the signal, we apply so-called exclusivity criteria, which rely on the kinematics of the di-$\PGt$ system and on the activity in the 
tracker part of the 
detector around the di-$\tau$ vertex. In particular, by selecting events with no track within 0.1\cm of the di-\PGt vertex, we can reduce the background contributions  
by about three orders of magnitude for a signal efficiency of around 50\%. Such a strategy has been used previously to identify similar photon-photon processes in $\Pp\Pp$ collisions~\cite{CMS:2011vma,CMS:2013hdf,CMS:2016rtz,CMS:2018uvs,ATLAS:2016lse,ATLAS:2020iwi}. 
Four different final states, corresponding to different decay modes of the $\PGt$ leptons, are studied to measure the signal strength: $\emu$, $\etau$, $\mutau$, and $\tautau$, where $\tauh$ denotes a
$\PGt$ lepton decaying hadronically. Events with two reconstructed muons are used to derive corrections to the simulated
samples. 
The visible mass of the $\PGt$ candidates, $\mvis$, computed as the invariant mass of the visible decay products of the two $\PGt$ leptons, 
is used as a biased estimator of $\mtt$ to extract the significance of the signal and set constraints on $\at$ and $\dt$.
Tabulated results are provided in the HEPData record for this analysis~\cite{hepdata}.

\section{The CMS detector}

The central feature of the CMS apparatus is a superconducting solenoid of 6\unit{m} internal diameter, providing a magnetic field of 3.8\unit{T}. Within the solenoid volume are a silicon pixel and strip tracker, a lead tungstate crystal electromagnetic calorimeter (ECAL), and a brass and scintillator hadron calorimeter (HCAL), each composed of a barrel and two endcap sections. Forward calorimeters extend the pseudorapidity coverage provided by the barrel and endcap detectors. Muons are measured in gas-ionization detectors embedded in the steel flux-return yoke outside the solenoid. A more detailed description of the CMS detector, together with a definition of the coordinate system used and the relevant kinematic variables, is presented in Ref.~\cite{CMS:2008xjf}. 

Events of interest are selected using a two-tiered trigger system. The first level, composed of custom hardware processors, uses information from the calorimeters and muon detectors to select events at a rate of around 100\unit{kHz} within a fixed latency of 4\mus~\cite{CMS:2020cmk}. The second level, known as the high-level trigger, consists of a farm of processors running a version of the full event reconstruction software optimized for fast processing, and reduces the event rate to around 1\unit{kHz} before data storage~\cite{CMS:2016ngn}.

\section{Signal and background modeling}

{\tolerance=800 Signal samples corresponding to the elastic photon-fusion production of a $\PGt$ lepton pair are generated using the \textsc{gamma-UPC} generator~\cite{Shao:2022cly}, which is implemented in the \MGvATNLO v3.5 code~\cite{Alwall:2014hca,Madgraph,Frederix:2012ps}, including the initial photon $\kt$ effects~\cite{Shao:2024dmk}. The \textsc{gamma-UPC} generator derives photon fluxes from electric dipole and charge form factors for protons and ions, and includes realistic hadronic survival probabilities for them.\par}
The photon fluxes used here are those derived from the elastic charge form factor of the proton. 
Samples generated with the \textsc{SuperChic} 4 generator~\cite{Harland-Lang:2020veo} are used as a cross-check, as described in Section~\ref{sec:signal}.

In this analysis, $\at$ and $\dt$ are parameterized in \MGvATNLO using the SMEFTsim package~\cite{Brivio:2017btx,Brivio:2020onw}, as suggested in Ref.~\cite{PhysRevD.102.113008}.
The corresponding BSM Lagrangian in the SM effective field theory (SMEFT) approach consists of two dimension-6 operators modifying $\at$ and $\dt$ at tree level:
\ifthenelse{\boolean{cms@external}}{
\begin{multline}
  \mathcal{L}_\text{BSM} = \frac{\CTB}{\Lambda^2} \overline{L}_L \sigma^{\mu\nu} \tau_R H B_{\mu\nu} \\ + \frac{\CTW}{\Lambda^2} \overline{L}_L \sigma^{\mu\nu} \tau_R \sigma^i H W^{i}_{\mu\nu} + \text{h.c.},
\end{multline}
}{
\begin{equation}
  \mathcal{L}_\text{BSM} = \frac{\CTB}{\Lambda^2} \overline{L}_L \sigma^{\mu\nu} \tau_R H B_{\mu\nu} + \frac{\CTW}{\Lambda^2} \overline{L}_L \sigma^{\mu\nu} \tau_R \sigma^i H W^{i}_{\mu\nu} + \text{h.c.},
\end{equation}
}
where $\Lambda$ is the scale of BSM physics, $L_L$ represents the left-handed $\PGt$ lepton doublet, $\tau_R$ corresponds to the right-handed 
$\PGt$ lepton, $H$ is an SU(2) scalar doublet, $B_{\mu\nu}$ is the weak hypercharge field, $W^{i}_{\mu\nu}$ represents the weak isospin field, and $\CTB$ and $\CTW$ are the Wilson coefficients. Using the SMEFTsim\_general\_alphaScheme\_UFO model~\cite{Brivio:2017btx,Brivio:2020onw}, 
after electroweak symmetry breaking, the $\PGt\PGt\PGg$ vertex is parameterized as:
\begin{equation}\label{eq:BSMgamma_mu}
  V_{\tau\tau\gamma} = ie\gamma^\mu - \frac{v\sqrt{2}}{\Lambda^2}\left[ \Re{\CTg} + \Im{\CTg}i\gamma_5\right]\sigma^{\mu\nu}q_\nu,
\end{equation} 
where $\CTg=\cos\theta_\mathrm{W} \CTB -  \sin\theta_\mathrm{W} \CTW$, $v\approx 246\GeV$ is the Higgs vacuum expectation value, and $\theta_\mathrm{W}$ is the weak mixing angle. 
Using Eqs.~(\ref{eq:gamma_mu}) and~(\ref{eq:BSMgamma_mu}), deviations from the SM predictions of $\at$ and $\dt$ can be expressed as:
\begin{equation}\label{eq:wilson}
  \delta \at  = \frac{2m_\PGt}{e}\frac{\sqrt{2}v}{\Lambda^2} \Re{\CTg}
\end{equation}
and
\begin{equation}
  \delta \dt  = \frac{\sqrt{2}v}{\Lambda^2} \Im{\CTg}.
\end{equation}

For simplicity and without loss of generality, since $\delta \at$ and $\delta \dt$ only depend on the linear combination of $\CTW$ and $\CTB$, we set $\CTW$ to 0 in the signal simulation. The coefficient $\CTB$ alone is modified to simulate different $\CTg$ values, which can be reinterpreted as linear combinations of $\CTW$ and $\CTB$. 
The BSM contributions to the $\PZ\to\PGt\PGt$ vertex cancel for $\CTW=-\tan\theta_\mathrm{W} \CTB$, but other values of $\CTW$ would result in variations in the Drell--Yan (DY) process prediction that are negligible with respect to the systematic uncertainties of this background, 
which are described in Section~\ref{sec:systematics}.
The simulation of different anomalous electromagnetic moments is performed using matrix element reweighting~\cite{Artoisenet:2010cn},
scanning over a range of $\CTB$ values.
In the simulation, we set $\Lambda$ to 2\TeV, but the results were verified to be independent of the scale of new physics.
Since the anomalous electromagnetic moments of the $\PGt$ lepton depend on a linear combination of $\CTW/\Lambda^2$ and $\CTB/\Lambda^2$, the analysis results can also be 
interpreted in this two-dimensional plane.
The form factor formalism of Eq.~(\ref{eq:gamma_mu}) and the SMEFT approach are equivalent assuming $q^2=0$.

The \textsc{gamma-UPC} generator is also used to simulate the elastic photon-fusion production of pairs of electrons, muons, and $\PW$ bosons, 
which are backgrounds in the signal region (SR).
Single- and double-dissociative processes are not simulated but estimated from data as described later.

The \MGvATNLO~2.6.5 event generator is used to generate events originating from the $\Zjets$ and $\Wjets$ 
processes, as well as from diboson production with two leptons and two neutrinos in the final state. These processes are simulated
 at next-to-leading order (NLO) with the FxFx jet matching and merging~\cite{Frederix:2012ps}. 
The \MGvATNLO generator is also used for the simulation of the minor contribution from other diboson production events, whereas 
\POWHEG 2.0~\cite{Nason:2004rx,Frixione:2007vw, Alioli:2010xd, Alioli:2010xa, Alioli:2008tz} is
 used for \ttbar and single top quark production. The generators are interfaced with
 \PYTHIA 8.240~\cite{Sjostrand:2014zea} to model parton showering and hadronization, the underlying event activity, as well as the decay of the
 $\PGt$ leptons. The \PYTHIA parameters affecting the description of the underlying event are set to
 the {CP5} tune~\cite{CMS:2019csb}.
The \textsc{NNPDF3.1} parton distribution function (PDF) set~\cite{Ball:2011uy,Ball:2013hta,Ball:2017nwa} at next-to-NLO accuracy is used for background simulations.
Additional $\Pp\Pp$ interactions per bunch crossing (pileup) are added to the simulated samples with a distribution that is chosen to match that observed in the data.
Generated events are processed through a \GEANTfour~\cite{Agostinelli:2002hh} simulation of the CMS detector.

\section{Object reconstruction}\label{sec:objects}

The global event reconstruction (also called particle-flow event reconstruction~\cite{CMS:2017yfk}) aims to reconstruct and identify each particle in an event, with an optimized combination of all subdetector information. In this process, the identification of the particle type (photon, electron, muon, charged hadron, or neutral hadron) plays an important role in the determination of the particle direction and energy. Photons are identified as ECAL energy clusters not linked to the extrapolation of any charged particle trajectory to the ECAL. Electrons are identified as a primary charged-particle track and potentially many ECAL energy clusters corresponding to this track extrapolation to the ECAL and to possible bremsstrahlung photons emitted along the way through the tracker material. Muons are identified as tracks in the central tracker consistent with either a track or several hits in the muon system and associated with calorimeter deposits compatible with the muon hypothesis. Charged hadrons are identified from the charged-particle tracks that are not associated with electrons or muons. Finally, neutral hadrons are identified as HCAL energy clusters not linked to any charged-hadron trajectory, or as a combined ECAL and HCAL energy excess with respect to the expected charged-hadron energy deposit. 

The electron momentum is estimated by combining the energy measurement in the ECAL with the momentum measurement in the tracker. The momentum resolution for electrons with transverse momentum $\pt \approx 45\GeV$ from $\PZ \to \Pe \Pe$ decays ranges 1.6--5.0\%. It is generally better in the barrel region than in the endcaps, and also depends on the bremsstrahlung energy emitted by the electron as it traverses the material in front of the ECAL~\cite{CMS:2020uim,CMS-DP-2020-021}. 

Muons are measured in the pseudorapidity range $\abs{\eta} < 2.4$, with detection planes made using three technologies: drift tubes, cathode strip chambers, and resistive-plate chambers. The efficiency to reconstruct and identify muons is greater than 96\%. Matching muons to tracks measured in the silicon tracker results in a relative \pt resolution of 1\% in the barrel and 3\% in the endcaps for muons with \pt up to 100\GeV~\cite{CMS:2018rym}. 

Hadronic \PGt lepton decays (\tauh) are reconstructed from jets, using the hadrons-plus-strips algorithm~\cite{CMS:2018jrd}, which 
combines 1 or 3 tracks with energy deposits in the calorimeters clustered as strips. Neutral pions are reconstructed as strips with dynamic size 
in $\eta$-$\phi$ (where $\phi$ is the azimuthal angle in radians) from reconstructed electrons and photons, where the strip size varies as a function of the $\pt$ of the electron 
or photon candidate. The \tauh candidates are reconstructed in four possible decay modes (DMs): a single charged hadron ($\mathrm{h}^\pm$), a charged hadron with at least one 
neutral pion ($\mathrm{h}^\pm+\PGpz$(s)), three charged hadrons with no neutral pion ($\mathrm{h}^\pm \mathrm{h}^\mp \mathrm{h}^\pm$), 
or 3 charged hadrons with at least one neutral pion ($\mathrm{h}^\pm \mathrm{h}^\mp \mathrm{h}^\pm + \PGpz$(s)).
To distinguish genuine \tauh decays from jets originating from the hadronization of quarks or gluons, and from electrons or muons, 
the \textsc{DeepTau} algorithm~\cite{CMS:2022prd} is used. Information from all individual reconstructed particles near the \tauh 
axis is combined with properties of the $\tauh$ candidate in the event.
The tight working point is used to separate $\tauh$ candidates from jets; its efficiency is about
65\% for $\tauh$ with $20<\pt<100\GeV$~\cite{CMS:2022prd}. The loosest working point, used in the background estimation procedure, has an efficiency
above 98\%.
The rate of a jet to be misidentified as \tauh by the \textsc{DeepTau} algorithm depends on the \pt and quark flavor of the jet. 
In simulated events from \PW boson production in association with jets, it has been estimated to be below the percent level for the tight working point. 
The misidentification rate for electrons (muons) is 2.60 (0.03)\% for a genuine \tauh identification efficiency of 80 ($>$99)\%. 

The missing transverse momentum vector \ptvecmiss is computed as the negative vector \pt sum of all the particle-flow candidates in an event, and its magnitude is denoted as \ptmiss~\cite{CMS:2019ctu}. The \ptvecmiss is modified to correct the energy scale of the reconstructed jets in the event.
Anomalous high-\ptmiss events can be due to a variety of reconstruction failures, detector malfunctions or noncollision backgrounds. Such events are rejected by event filters that are designed to identify more than 85--90\% of the spurious high-\ptmiss events with a mistagging rate less than 0.1\%~\cite{CMS:2019ctu}. 

Tracks with $\pt>0.5\GeV$ and $\abs{\eta}<2.5$ are used in this analysis to measure the hadronic activity in the event. 
The silicon tracker used in 2016 measured charged particles within the range $\abs{\eta} < 2.5$. For nonisolated particles of $1 < \pt < 10\GeV$ and $\abs{\eta} < 1.4$, the track resolutions were typically 1.5\% in \pt and 25--90 (45--150)\mum in the transverse (longitudinal) impact parameter~\cite{CMS:2014pgm}. At the start of 2017, a new pixel detector was installed~\cite{Phase1Pixel}; the upgraded tracker measured particles up to $\abs{\eta} < 3.0$ with typical resolutions of 1.5\% in \pt and 20--75\mum in the transverse impact parameter~\cite{DP-2020-049} for nonisolated particles of $1 < \pt < 10\GeV$.

During the 2016--2017 data taking, a gradual shift in the timing of the inputs of the ECAL first-level trigger in the region at $\abs{\eta} > 2.0$ caused a specific trigger inefficiency~\cite{EGM-18-002}. For events containing an electron (a jet) with \pt larger than $\approx$50\GeV ($\approx$100\GeV), the efficiency loss is $\approx$10--20\% in the region $2.5 < \abs{\eta} < 3.0$, depending on \pt, $\eta$, and time. Correction factors were computed from data and applied to the acceptance evaluated by simulation.

\section{Event selection}\label{sec:selection}

Events in the SR are selected in the $\emu$, $\etau$, $\mutau$, and $\tautau$ final states, which account for 94\% 
of the possible final states with a pair of $\PGt$ leptons. The $\Pe\Pe$ and $\mumu$ final states have the lowest branching 
fractions for di$-\Pgt$ events and suffer from large background contributions from the photon-fusion production 
of electron and muon pairs, $\ggtoee$ and $\ggtomumu$. Therefore, they are not considered as part of the 
signal. However, the $\mumu$ final state, which, in comparison to the $\Pe\Pe$ final state, benefits from 
lower \pt thresholds at trigger level and higher lepton reconstruction and identification efficiencies, 
is used to derive corrections to: 
\begin{itemize}
  \item the pileup track density in all simulations, described in Section~\ref{sec:ntracksPU};
  \item the hard scattering track multiplicity in the DY and diboson simulations, described in Section~\ref{sec:ntracksHS};
  \item the elastic photon-fusion simulations to include dissociative contributions, described in Section~\ref{sec:signal}; 
  \item the acoplanarity distribution in the DY simulation, described in Section~\ref{sec:dy}. 
\end{itemize}

In the $\emu$ final state, events are recorded with a combination of triggers requiring an electron and a muon, with a \pt threshold 
of 23\GeV for the leading lepton, and 8 (12)\GeV for the subleading muon (electron). 
Offline, the leading (subleading) lepton is required to have $\pt>24$ (15)\GeV. The subleading lepton \pt threshold 
is chosen to reduce background processes with misidentified jets. 
In the $\etau$ ($\mutau$) final states, events are recorded with single-electron (single-muon) triggers with $\pt$ thresholds ranging between 
25 and 32 (24 and 27)\GeV depending on the data-taking year. Additional events with \Pe or \PGm \pt as low as 24 (19)\GeV and $\abs{\eta}<2.1$ 
can be selected using 
triggers requiring also the presence of a $\tauh$ candidate with $\pt>20$ (27)\GeV and $\abs{\eta}<2.1$. In these 
two final states, the $\tauh$ \pt is required to be above 30\GeV offline and the background 
from $\Wjets$ is reduced by requiring the transverse mass $\mT$ between the \Pe or \PGm \pt and $\ptvecmiss$ to be less than 75\GeV.
The events in the $\tautau$ final state are selected with triggers requiring the presence of two loosely isolated $\tauh$ candidates with 
$\pt>35$--40\GeV, depending on the data-taking year, and $\abs{\eta}<2.1$. The offline thresholds are $\pt>40\GeV$ and $\abs{\eta}<2.1$ for all the data-taking periods. 
A trigger requiring a single muon with $\pt>24$--27\GeV is used to select events in the $\mumu$ final state. Offline, the leading 
(subleading) muon 
is required to have $\pt>26$--29 (10)\GeV and $\abs{\eta}<2.4$. The invariant mass of the two muons, $\mmumu$, must be greater than 50\GeV.

In all final states, the two objects are required to be well-identified and isolated, to have opposite sign (OS) charge, 
to be separated from each other by $\Delta R = \sqrt{\smash[b]{(\Delta \eta)^2 + (\Delta \phi)^2}} > 0.5$, 
and to satisfy $\abs{d_z(\Pell,\Pell')}<0.1\cm$, where $\abs{d_z(\Pell,\Pell')}$ is the difference in the longitudinal impact parameters of the two objects. 
In each final state, we veto events that have additional identified and
isolated electrons or muons, such that one event can enter at most one final state. 

The criteria listed above and summarized in Table~\ref{tab:selection} constitute the baseline selection. Additional requirements are applied to define the SR or the control regions (CRs) used to derive corrections. Additional requirements increasing
the signal-to-background ratio in the SR using the characteristics of the exclusive process signature 
are described in the following paragraphs. 

\begin{table*}
\topcaption{Baseline selection criteria used in the different final states. The electrons, muons, and $\tauh$ are required to be well identified and isolated. The \pt and pseudorapidity ranges correspond to different sets of triggers, and different data-taking periods.}\label{tab:selection}
\renewcommand{\arraystretch}{1.2}
\centering
\begin{tabular}{llllll}
 & $\emu$ & $\etau$ & $\mutau$ & $\tautau$ & $\mumu$ \\
\hline
	$\pt^{\Pe}$ (\GeVns) & $>15/24$  & $>25$--33 & \NA & \NA & \NA \\
$\abs{\eta^{\Pe}}$ & $<2.5$ & $<2.1$--2.5 & \NA & \NA & \NA \\
	$\pt^{\PGm}$ (\GeVns) & $>24/15$ & \NA & $>21$--29 & \NA & $>26$--29/10 \\
$\abs{\eta^{\PGm}}$ & $<2.4$ & \NA & $<2.1$--2.4 & \NA & $<2.4$ \\
	$\pttauh$ (\GeVns) & \NA & $>30$--35 & $>30$--32 & $>40$ & \NA \\
$\abs{\eta^{\tauh}}$ & \NA & $<2.1$--2.3 & $<2.1$--2.3 & $<2.1$ & \NA \\
	$\mmumu$ (\GeVns) & \NA & \NA & \NA & \NA & $>50$ \\
OS & yes & yes & yes & yes & yes \\
	$\abs{d_z(\Pell,\Pell')}$ (\cmns) & $<0.1$ & $<0.1$ & $<0.1$ & $<0.1$ & $<0.1$ \\ 
$\Delta R(\Pell,\Pell')$ & $>0.5$ & $>0.5$ & $>0.5$ & $>0.5$ & $>0.5$ \\
	$\mT(\Pe/\PGm \pt,\ptvecmiss)$ (\GeVns)& \NA & $<75$ & $<75$ & \NA & \NA \\
\end{tabular}
\end{table*}

The acoplanarity, defined as:
\begin{equation}
  A=1-\abs{\Delta\phi(\Pell,\Pell')}/\pi,
\end{equation}
is typically small for events from the elastic photon-fusion production, for which the leptons are produced back-to-back. In the case of $\PGt$ leptons, 
$A$ is computed using the visible decay products, which are closely aligned with the original $\PGt$ lepton before decay, especially at high 
\pt. In the SR, 
the events are required to have $A<0.015$. This requirement has a signal efficiency ranging from 95\% in the $\emu$ final state to 
$>99\%$ in the $\tautau$ final state. 

Elastic signal events are also characterized by a low track multiplicity ($\ntracks$) at the dilepton vertex. The dilepton vertex 
position along the $z$ axis is calculated as the average $z$ of the two objects. The variable $\ntracks$ is estimated by 
counting the number of tracks that have a $z$ coordinate within a 0.1\cm wide window centered around the dilepton vertex, removing 
the tracks that were used to build the lepton candidates. The window size is chosen such that about 50\% of the
signal events have $\ntracks=0$. This fraction is larger far from the beamspot position ($\zBS$), where the contribution from 
pileup tracks is less important. The SRs are defined with $\ntracks=0$ or 1. About 25\% of the 
signal events have $\ntracks=1$. 

A low pileup activity in the tracker, corresponding to low $\ntracks$ values at the interaction vertex, impacts the identification of electrons, muons, and $\tauh$ candidates, typically leading to a higher identification and isolation efficiency. 
Scale factors are computed to match the simulated efficiencies to the observed efficiencies for objects surrounded by a low track activity at their vertex.

\section{Corrections to the simulated track multiplicity}\label{sec:corrections}

The simulations are corrected to accurately describe $\ntracks$.
We derive corrections for two sources of tracks:
tracks originating from the pileup (Section~\ref{sec:ntracksPU}) and 
tracks originating from the hard scattering and underlying event activity (Section~\ref{sec:ntracksHS}).  
They are derived from the dimuon CR and applied to simulations in the SRs.

\subsection{Tracks originating from pileup}\label{sec:ntracksPU}

While the pileup multiplicity in simulation is reweighted to match that observed in the data, the pileup track density along the $z$ axis 
in simulation needs to be corrected to better describe $\ntracks$ at the dilepton vertex. 
First, the simulations are corrected, independently for each data-taking period, such that the $\zBS$ 
and beamspot width ($\sBS$) distributions are similar to the observed ones. In data collected between 2016--2018, $\zBS$ ($\sBS$) is 
approximately in the range between -0.5 and +1.2\cm (3.1 and 4.2\cm). The pileup tracks are redistributed 
accordingly along the $z$ axis, as explained below. Simulated events originally all have 
$\sBS=3.5\cm$, whereas the observed $\sBS$ distribution is spread between about 3.0 and 3.8\cm. Simulated events 
are assigned randomly corrected beamspot properties, $\zBS^\text{corr}$ and $\sBS^\text{corr}$, following the 
corresponding profiles in data. The $z$ positions of pileup tracks, 
identified as such using generator-level information, are corrected for beamspot effects:
\begin{equation}
  z^\text{corr} = \zBS^\text{corr} + \frac{\sBS^\text{corr}}{\sBS^\text{sim}}\left(z-\zBS^\text{sim}\right).
\end{equation} 

\begin{figure*}[ht]
  \centering
  \includegraphics[width=0.95\textwidth]{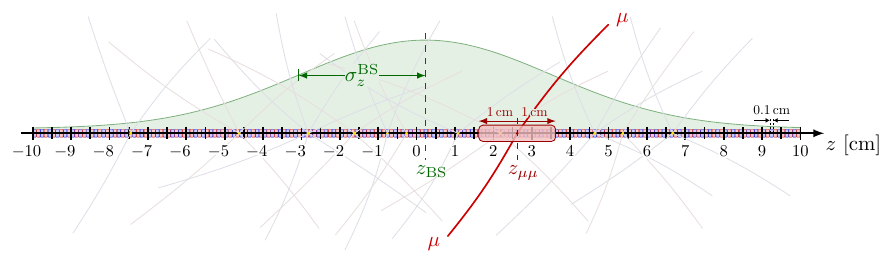}
  \caption{Schematic view of the 0.1\cm wide windows probed along the $z$ axis to derive corrections to the pileup track density in simulation. Windows within 1\cm from the dimuon vertex, illustrated with the red box, are discarded so as not to count tracks from the hard-scattering interaction. The green curve indicates the probability distribution of z-coordinates for PU vertices in the beamspot.}\label{fig:beamspot}
\end{figure*}

\begin{figure}
  \centering
  \includegraphics[width=0.42\textwidth]{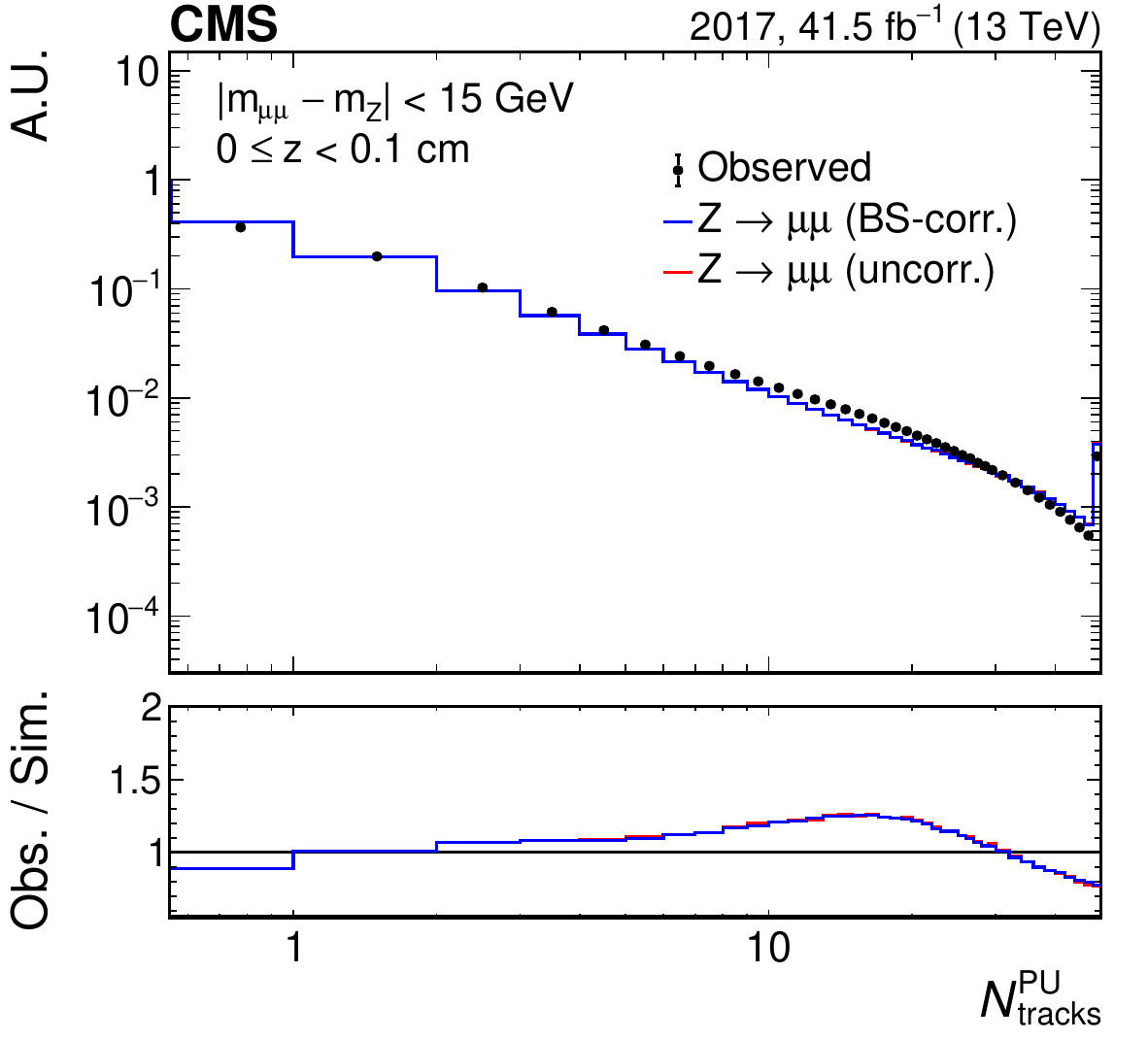}
  \includegraphics[width=0.42\textwidth]{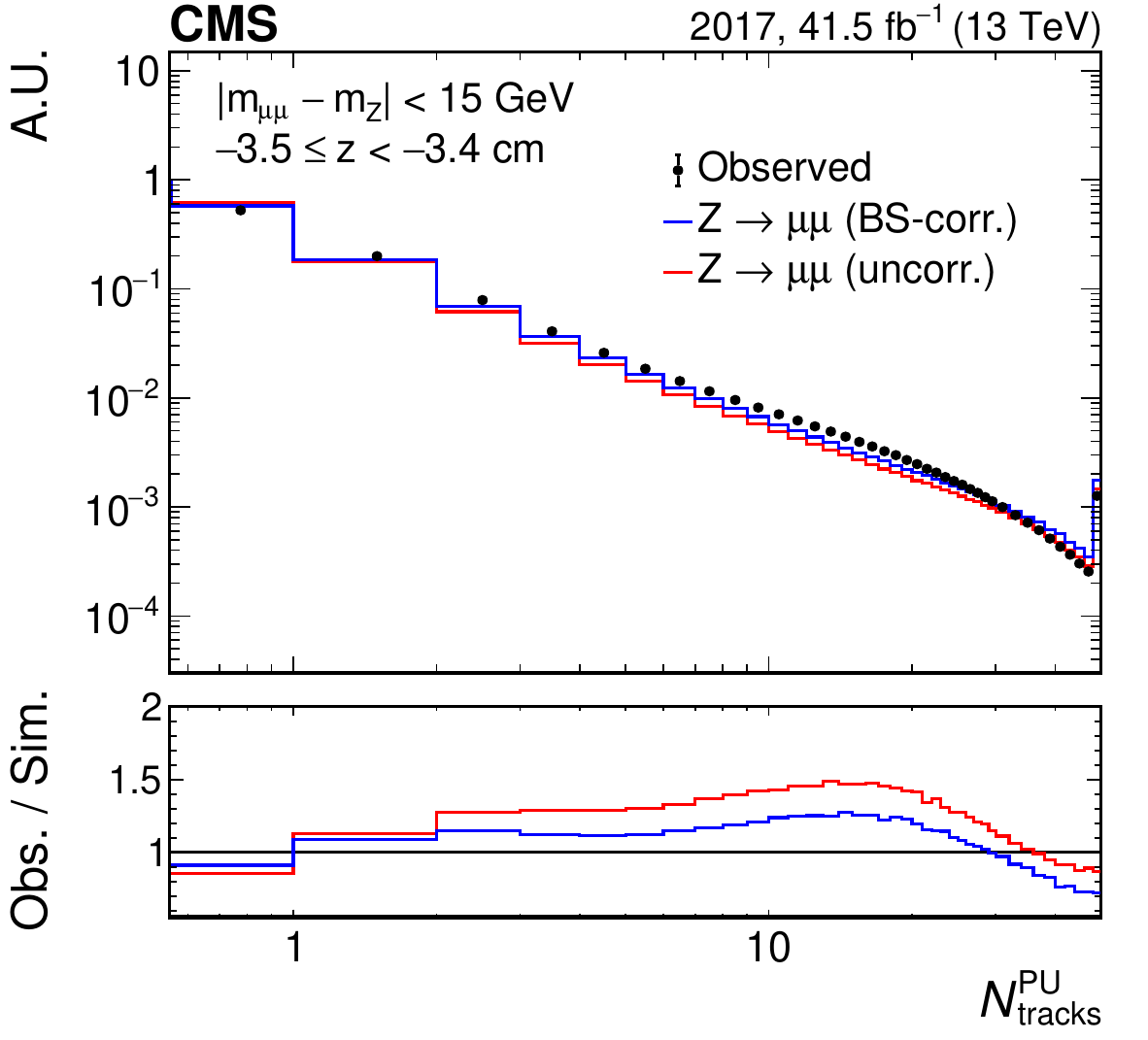}\\
  \includegraphics[width=0.42\textwidth]{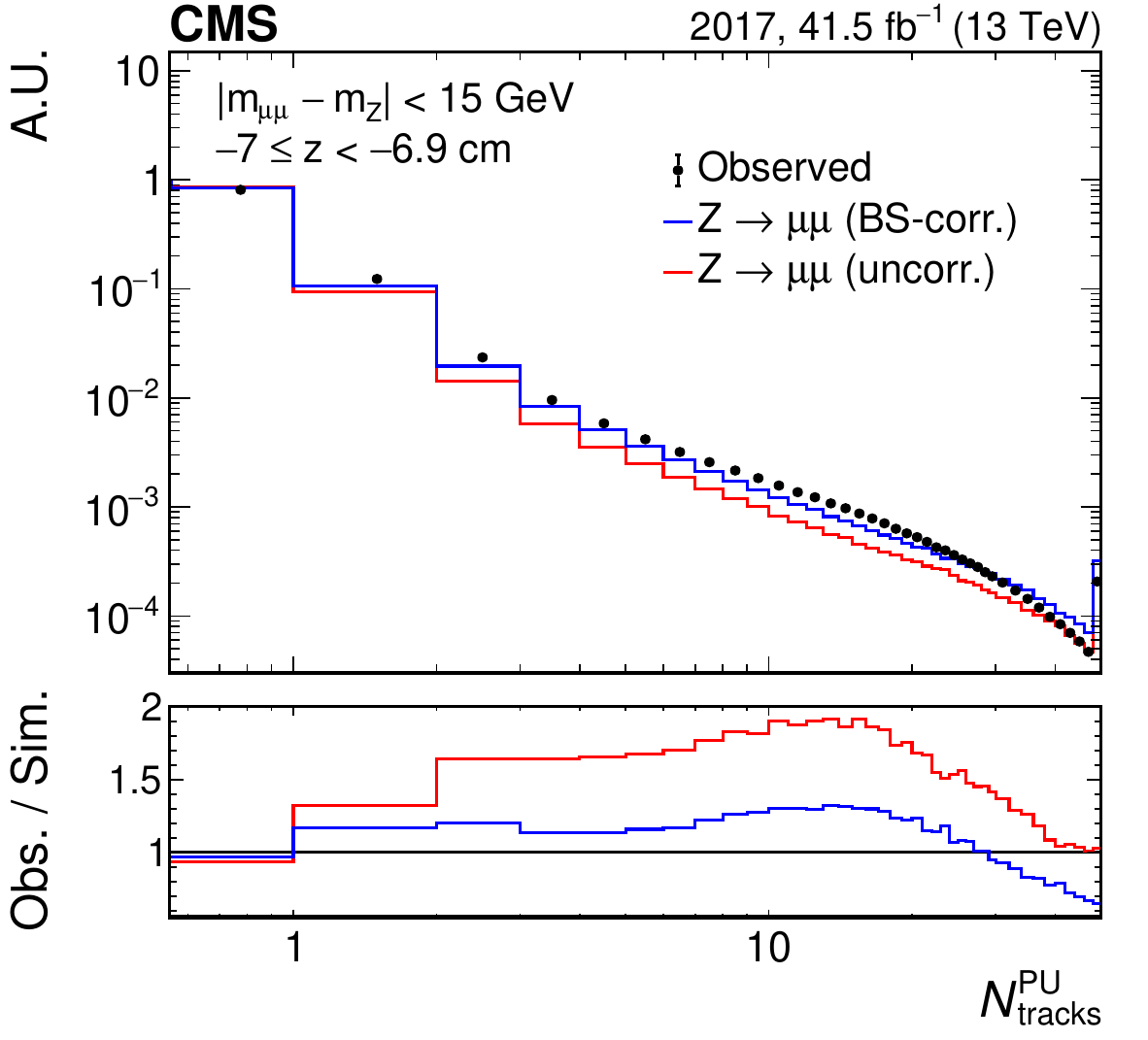}
	\caption{Distribution of $\ntracksPU$ in windows of 0.1\cm width along the $z$ axis for the observed events (black), uncorrected simulation (red), and beamspot-corrected simulation (blue) for data collected in 2017. The windows shown here are located at the beamspot center (\cmsUpperLeft), and one (\cmsUpperRight) or two (lower) beamspot widths away from the center. The ratio of beamspot-corrected simulation to observation (lower plots) is taken as a residual correction to the simulations. The last bin includes the overflow. Similar distributions and corrections are derived independently for the other data-taking periods.}\label{fig:corr_npu}
\end{figure}

A residual correction to the pileup track density is derived in a $\PZ\to\PGm\PGm$ CR, 
defined with the selection criteria for the $\mumu$ final state detailed in Section~\ref{sec:selection}, with the additional requirement that 
$\mmumu$ is within 15\GeV of the $\PZ$ boson mass, $\mZ$. The track multiplicity is counted in data and in simulation in windows 
of 0.1\cm along the $z$ axis, for windows centered at least 1\cm away from the reconstructed dimuon vertex, as illustrated in 
Fig.~\ref{fig:beamspot}. This last requirement 
ensures that all tracks in the windows come from pileup and not from the hard-scattering interaction. Weights are derived as functions of the pileup 
track multiplicity and window $z$ position so as to match the $\ntracks$ distribution of the observed events. 
Distributions of the number of pileup tracks,
$\ntracksPU$, are shown for the measured events, and for the uncorrected and beamspot-corrected simulations in Fig.~\ref{fig:corr_npu}, 
for windows with different $z$ positions. About one third of the windows situated at the beamspot center do not have any pileup track. This fraction 
reaches about 50 (80)\% one (two) beamspot width(s) away from the beamspot center. 
The correction to the simulated event weight is in the range 0.89--1.00 (0.93--1.16) for 
$\ntracksPU=0$ ($\ntracksPU=1$) over the whole $z$ axis and data-taking periods. It is evaluated at the z position of the selected dilepton candidate. 

\subsection{Tracks originating from the hard scattering and underlying event activity}\label{sec:ntracksHS}

The only major background in the SR estimated using simulation is the DY production of a lepton pair. The number of tracks 
coming from the hard scattering interaction is not well described in the simulation, and corrections are derived from the same $\PZ\to\PGm\PGm$ CR, 
in which $\mmumu$ is required to be within 15\GeV of $\mZ$, by comparing 
the number of tracks around the reconstructed dimuon vertex between data and simulation. 

\begin{figure}[ht]
\centering
  \includegraphics[width=0.49\textwidth]{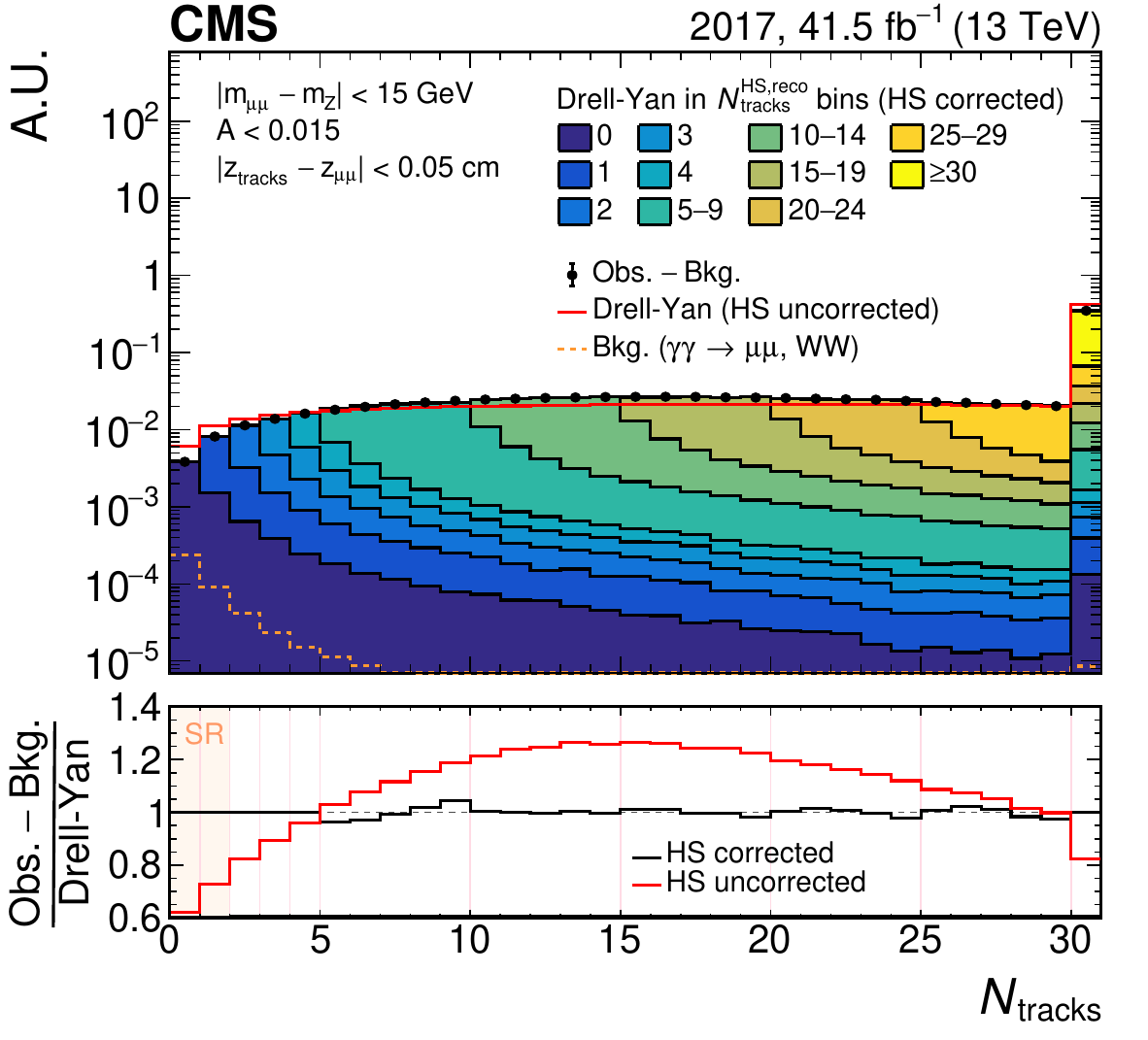}
  \caption{
Distribution of the number of reconstructed tracks in a 0.1\cm wide window in the $z$ direction, 
centered on the dimuon reconstructed vertex, for acoplanarity $A<0.015$, in data collected in 2017.
The DY simulation is split into several components based on the number of reconstructed tracks originating from the hard interaction.
The red line shows the simulation before the correction.
The black points show the observed data after subtracting the expected background contribution from the $\ggtomumu$ and $\ggtoww$ processes (dashed orange line).
The last bin includes the overflow. Similar distributions and corrections are derived independently for the other data-taking periods.
The ratios between the observed data, from which the exclusive background contributions have been subtracted, and the DY prediction before (red) and after the corrections (black), are shown in the lower panel.
The region with the selection requirement $\ntracks=0$ or 1 used in the SR is highlighted with the orange shaded area in the lower panel.
  }\label{fig:corr_npu_nhs}
\end{figure}

As shown in Fig.~\ref{fig:corr_npu_nhs}, 
the simulation is split into several components depending on the number of tracks with $\pt>0.5\GeV$ and $\abs{\eta}<2.5$, identified as originating from the hard scattering interaction 
using generator-level information, and not matched to the selected muons.
The total number of reconstructed tracks, excluding the two muon tracks, is the 
sum of these tracks and those associated with pileup interactions: $\ntracks=\ntracksHS+\ntracksPU$. Events from the exclusive photon-fusion 
production of a pair of muons or $\PW$ bosons, which contribute significantly only at low $\ntracks$, are estimated from simulation and 
subtracted.

The pileup track density is first corrected with the weights described 
in Section~\ref{sec:ntracksPU}, whereas the reweighting factors for each DY component with a given number of $\ntracksHS$ 
is determined iteratively by matching the simulation to the observed data, starting from events with $\ntracks=0$, to which only the 
simulated component with $\ntracksHS=0$ contributes. The corrections are determined separately for the different data-taking years. 
We checked that the requirement for $\mmumu$ be compatible with $\mZ$ introduces a bias for events with large dilepton mass, which is negligible with respect to other systematic uncertainties described in Section~\ref{sec:systematics}. 
The event weight correction is as low as $0.61\pm0.04$ ($0.76\pm0.04$) for $\ntracksHS=0$ ($\ntracksHS=1$). 
This correction is also applied to the minor diboson background 
with 2 leptons and 2 neutrinos in the final state, which has the same generator settings as the DY simulation. 

\section{Signal estimation}\label{sec:signal}

The signal is modeled using events of elastic $\Pp\Pp$ scattering. The semidissociative and fully dissociative production modes are much less likely to 
satisfy the acoplanarity and $\ntracks$ requirement, but they contribute a nonnegligible number of events to the SRs because of their 
larger cross sections. 

\begin{figure}
  \centering
  \includegraphics[width=0.49\textwidth]{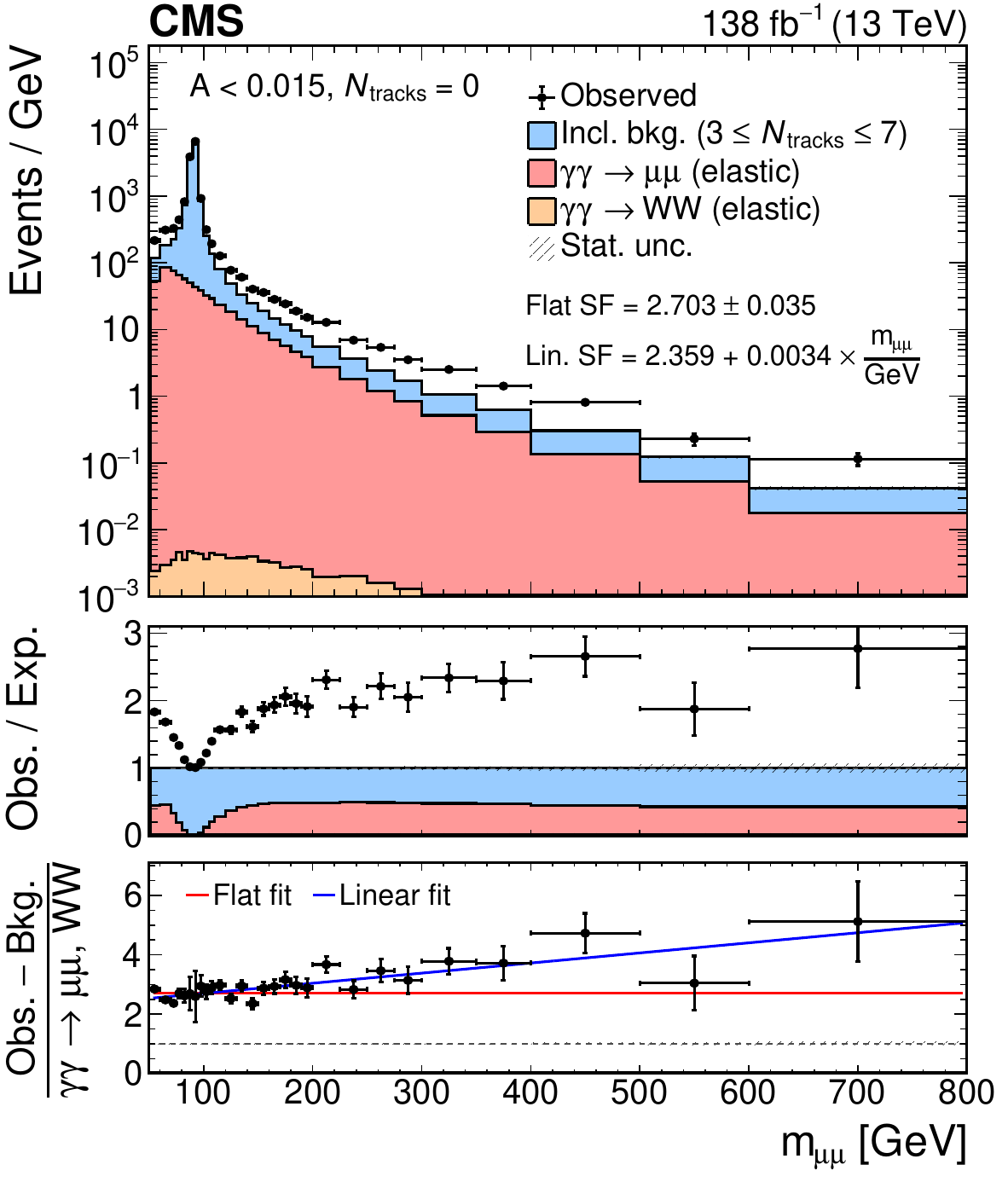}
  \includegraphics[width=0.49\textwidth]{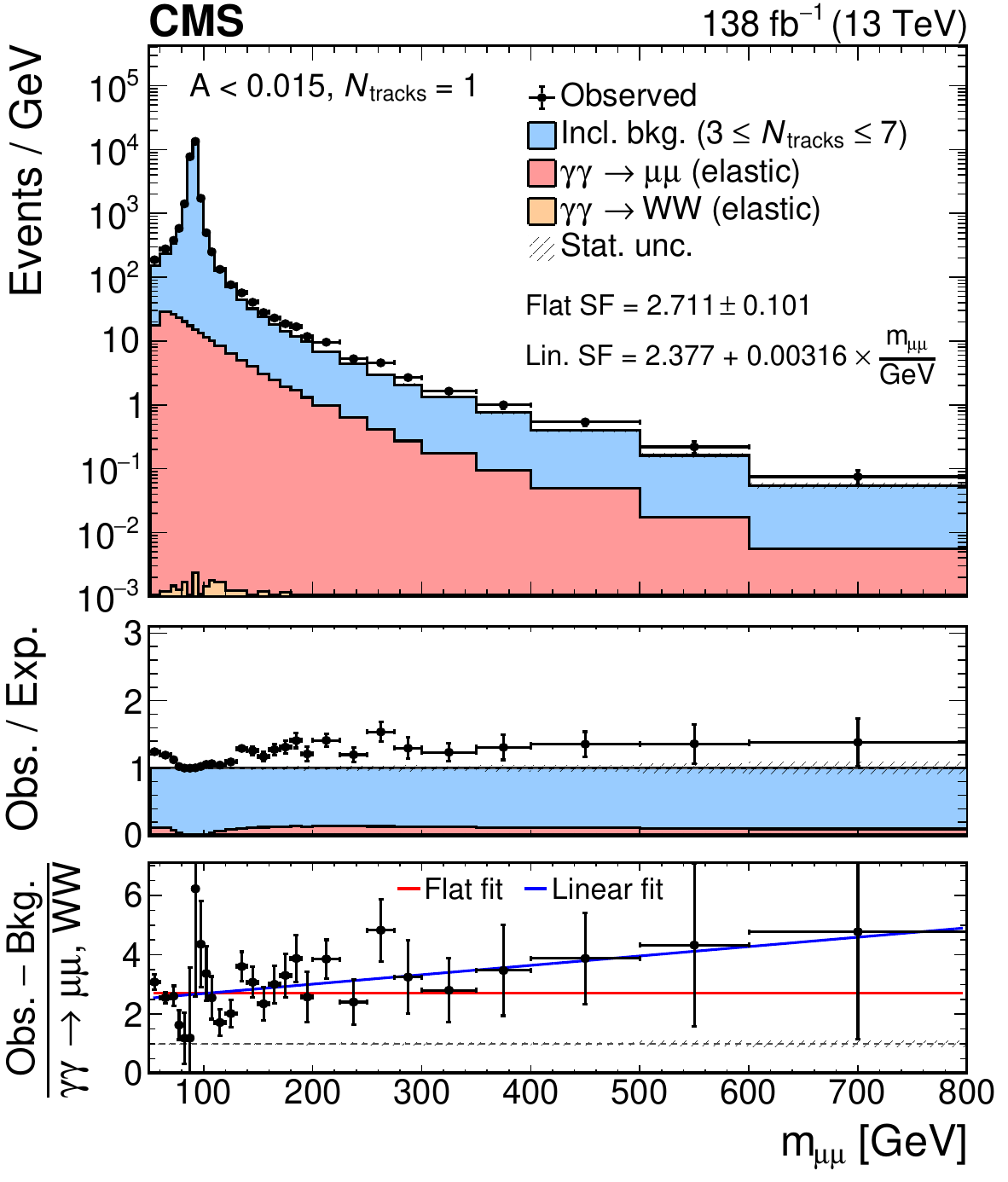}
  \caption{
Measurement of the scale factor for the elastic exclusive signal in $\mumu$ events for $\ntracks=0$ (\cmsLeft) or 1 (\cmsRight), and $A<0.015$.
The shape of the inclusive background (blue) is estimated from the observed data in the $3\leq\ntracks\leq7$ sideband, and rescaled to fit the observed data in $75<\mmumu<105\GeV$.
The scale factor is fitted in the lower ratio panel with  constant (red) and linear (blue) functions. The vertical error bars indicate the statistical uncertainty in the number of observed events.
  }\label{fig:rescaling}
\end{figure}

The prediction from the elastic photon-fusion production is rescaled to account for the semidissociative and fully dissociative contributions. 
The scaling factor comes from data observed in a $\mumu$ CR, defined with the baseline selection summarized in Table~\ref{tab:selection} and $\ntracks=0$ or 1 and $A<0.015$ as additional requirements. 
To perform this measurement, the inclusive backgrounds, dominated by the DY production of a muon pair, are 
estimated together from observed events. Their $\mmumu$ distribution is taken from observed events with $3<\ntracks<7$. These thresholds ensure that 
the photon-fusion contributions are negligible, while preserving the $\mmumu$
distribution. The compatibility of the $\mmumu$
distributions in inclusive events with $\ntracks<2$ and $3<\ntracks<7$ was verified in simulation. 
The $\mmumu$ distribution is scaled so that the inclusive background normalization 
in the range $\abs{\mmumu-\mZ}<15\GeV$ matches the event count observed in the same range. The $\ggtomumu$ prediction for the elastic production  
is then rescaled to match 
the difference between the experimental data and the inclusive backgrounds away from the $\PZ$ peak region, as shown in Fig.~\ref{fig:rescaling}. 

Assuming no dependence on the 
dilepton mass, $\mll$, the scaling factor to include the dissociative contributions is $2.70\pm0.04$ ($2.71\pm0.10$) for $\ntracks=0$ (1), where 
the uncertainty is statistical only. 
This is compatible with the value predicted by the \textsc{SuperChic} generator, using the same acoplanarity requirement $A<0.015$. The scale factor is also 
fitted as a linear function of $\mmumu$, which demonstrates a better $\chi^2$, and, in that case, is equal to $2.36+0.0034~\mmumu/\GeVns$ ($2.38+0.0032~\mmumu/\GeVns$) for $\ntracks=0$ (1). The fit probability is also better with a linear function for the predictions of the \textsc{SuperChic} generator, and a comparable parameterization is obtained.
A linear dependency of the scaling factor with $\mll$ is assumed as a nominal correction in the SRs, whereas the flat value is considered as a systematic 
uncertainty as detailed in Section~\ref{sec:systematics}. Elastic events in the SR, coming from the $\ggtotautau/\Pe\Pe/\PGm\PGm/\PW\PW$ simulations, are all rescaled with this 
correction as a function of the dilepton or diboson mass, to include semidissociative and fully dissociative contributions. 

\section{Background estimation}

In the SR, the dominant backgrounds are the DY production of a lepton pair and processes with a jet misidentified as an electron, 
a muon, or a $\tauh$ candidate, collectively denoted as ``jet mis-ID background''. This background is dominantly composed of 
$\Wjets$ and quantum chromodynamics (QCD) multijet events. 
The photon-fusion production of a pair of $\PW$ bosons is reduced with the acoplanarity requirement, and is significant only in the 
$\emu$ final state, where its expected contribution is about 4\% of the exclusive di-\PGt production. The photon-fusion 
production of an electron (muon) pair is about 50 (4)\% of the expected signal contribution in the $\etau$ ($\mutau$) final state. When there is no 
exclusivity requirement relying on $A$ or $\ntracks$, the $\ttbar$, inclusive diboson, and single top quark processes also contribute. 

\subsection{Drell--Yan background}\label{sec:dy}

\begin{figure}
  \centering
  \includegraphics[width=0.49\textwidth]{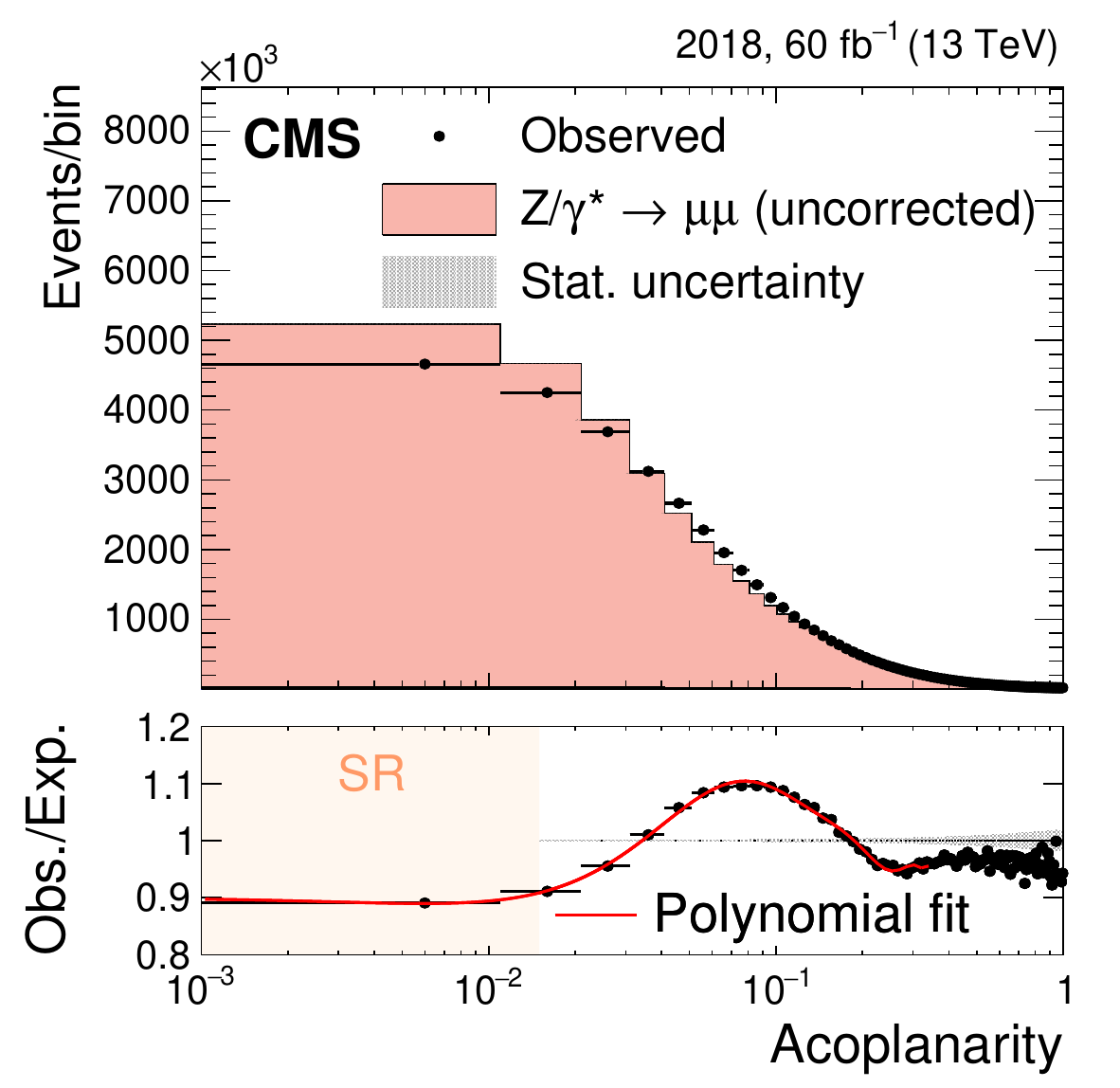}
  \caption{
	  Acoplanarity distribution for the observed events and in DY simulation before correction, in the 2018 data-taking period. The background prediction is normalized to match the observed yield and only the statistical uncertainty is shown. The data-to-simulation ratio is fitted with a polynomial to obtain the correction. The selection criterion $A<0.015$ used in the SR is highlighted with the orange shaded area in the lower panel.
  }
  \label{fig:aco}
\end{figure}

The DY background is estimated using simulation. The corrections described in Section~\ref{sec:corrections} are applied to 
improve the description of $\ntracksPU$ and $\ntracksHS$. 
The acoplanarity is observed to be mismodeled by simulation in a $\mumu$ CR, defined with the baseline selection criteria from Section~\ref{sec:selection} and 
further enriched in DY events by requiring 
$\abs{\mmumu-\mZ}<15\GeV$. An acoplanarity correction, measured in two-dimensional bins of the \pt of both muons, is derived in this CR by matching 
the simulation to the observed data. The correction is obtained by fitting the ratio of data-to-simulation with a polynomial for $A<0.35$. 
For illustrative purposes, the distributions and correction obtained for all muon \pt bins merged together are shown 
in Fig.~\ref{fig:aco}, using data collected in 2018. This acoplanarity  
correction is applied to simulated DY events in the SR on the basis of the generated $\PGt$ lepton \pt before their 
decays. 

The size of the DY simulation in the final state of a pair of electrons, muons, or $\PGt$ leptons is limited, 
and large statistical uncertainties arise when requiring $\ntracks=0$ or 1. 
To reduce this effect, the DY contribution in the SR is estimated from simulated events with $\ntracks<10$, which constitute about 10\% of the total 
number of DY events. The upper bound ensures 
a similar event topology as in the SR, ensuring that no bias is introduced to the $\mtt$ distributions, as verified in simulation. The distribution derived with $\ntracks<10$ is 
scaled to predict the DY events with $\ntracks=0$ or 1. The scale factors are derived in the $\emu$ final state with a 
simulation with a large number of $\PZ/\PGg*\to\PGt\PGt$ events with only leptonic $\PGt$ decays. All the corrections described in Section~\ref{sec:corrections} 
are applied to predict the DY yield in the $\emu$ final state with $\ntracks=0$, 1, or $<10$. The scale factors are equal to $(2.48\pm0.05)\%$ and 
$(5.10\pm0.07)\%$, for $\ntracks=0$ and 1, respectively. In the $\emu$ final state, the large simulation with leptonic $\PGt$ decays 
is used instead of reweighting events with $\ntracks<10$. 

\subsection{Backgrounds with misidentified jets in the \texorpdfstring{$e\tau_h$}{etauh} and \texorpdfstring{$\mu\tau_h$}{mutauh} final states}

The mis-ID background with a jet misidentified as a $\tauh$ candidate is estimated from the experiment. 
In the $\etau$ and $\mutau$ final states, the dominant processes contributing to the mis-ID background are $\Wjets$ 
and QCD multijet events. The jet misidentification factors (MFs) depend on the process because of different topologies and jet flavors, 
and they are therefore measured in two different CRs. A CR enriched in $\Wjets$ events is built by requiring $\mT>75\GeV$, whereas a QCD-enriched 
CR is obtained by requiring the two objects to have same-sign (SS) charge. There is no requirement on $\ntracks$ at this stage. 
In these CRs, the MFs are measured for each data-taking period,
separately for each $\tauh$ DM ($\DMtauh$), by taking 
the ratio of events in which the $\tauh$ candidate passes the SR $\tauh$ identification criteria, to events in which the $\tauh$ candidate
fails these criteria but passes looser ones. Events where the $\tauh$ candidate is genuine or corresponds to a misidentified electron or muon 
are evaluated using simulation and subtracted from the CR. The MFs are fitted as functions of the $\tauh$ \pt and range from about 0.03 for the 
$\mathrm{h}^{\pm}\mathrm{h}^{\mp}\mathrm{h}^{\pm}+\PGpz$(s) DM, to about 0.40 for the $\mathrm{h}^{\pm}$ DM. The MFs are measured in events recorded with 
single-electron or -muon triggers. The triggers that select $\tauh$ candidates in addition to electrons or muons, called ``cross-triggers'', apply loose isolation requirements to the $\tauh$ candidates 
at trigger level, such that the MFs are modified. The MF measured in events selected with cross-triggers is measured to be 1.7 times higher than the average MF in 
events selected with single-lepton triggers, and this factor is applied as a multiplicative correction to the MFs for events recorded with a cross-trigger. 

\begin{figure}
  \centering
  \includegraphics[width=0.49\textwidth]{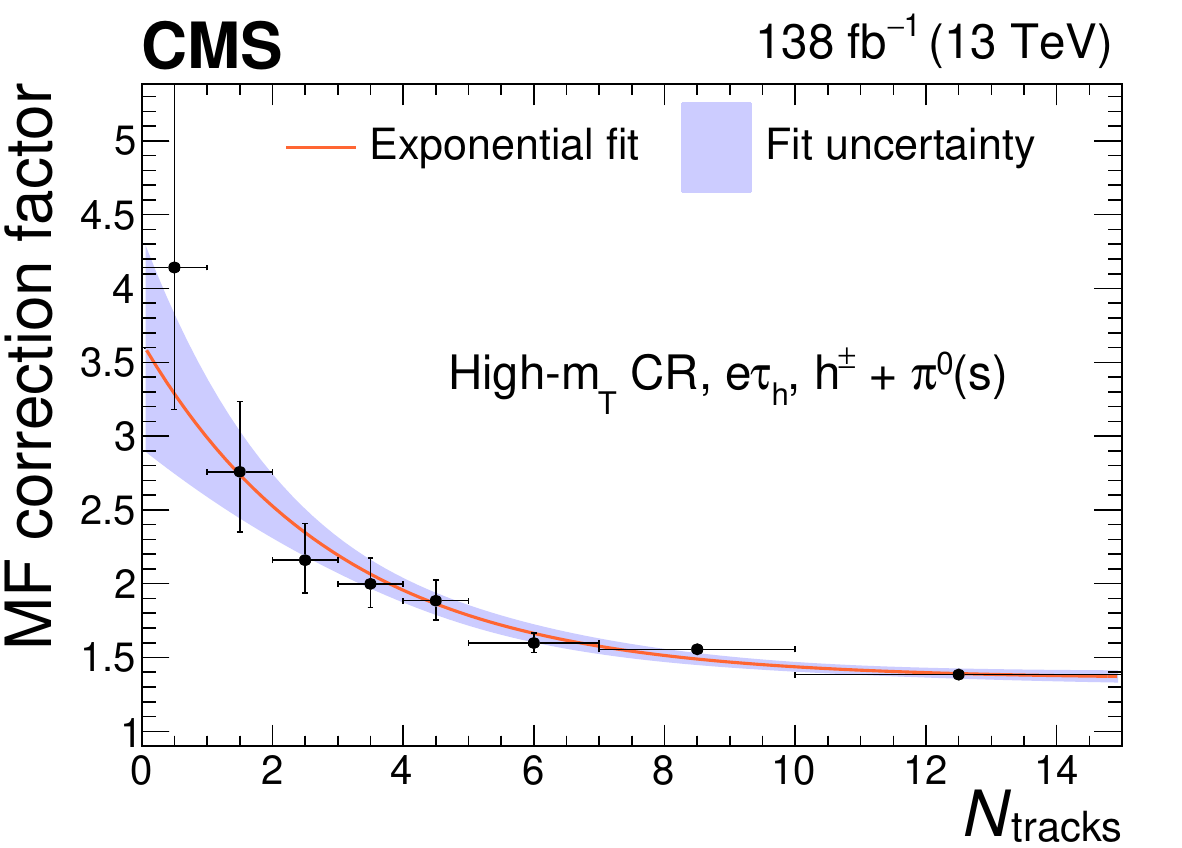}
  \includegraphics[width=0.49\textwidth]{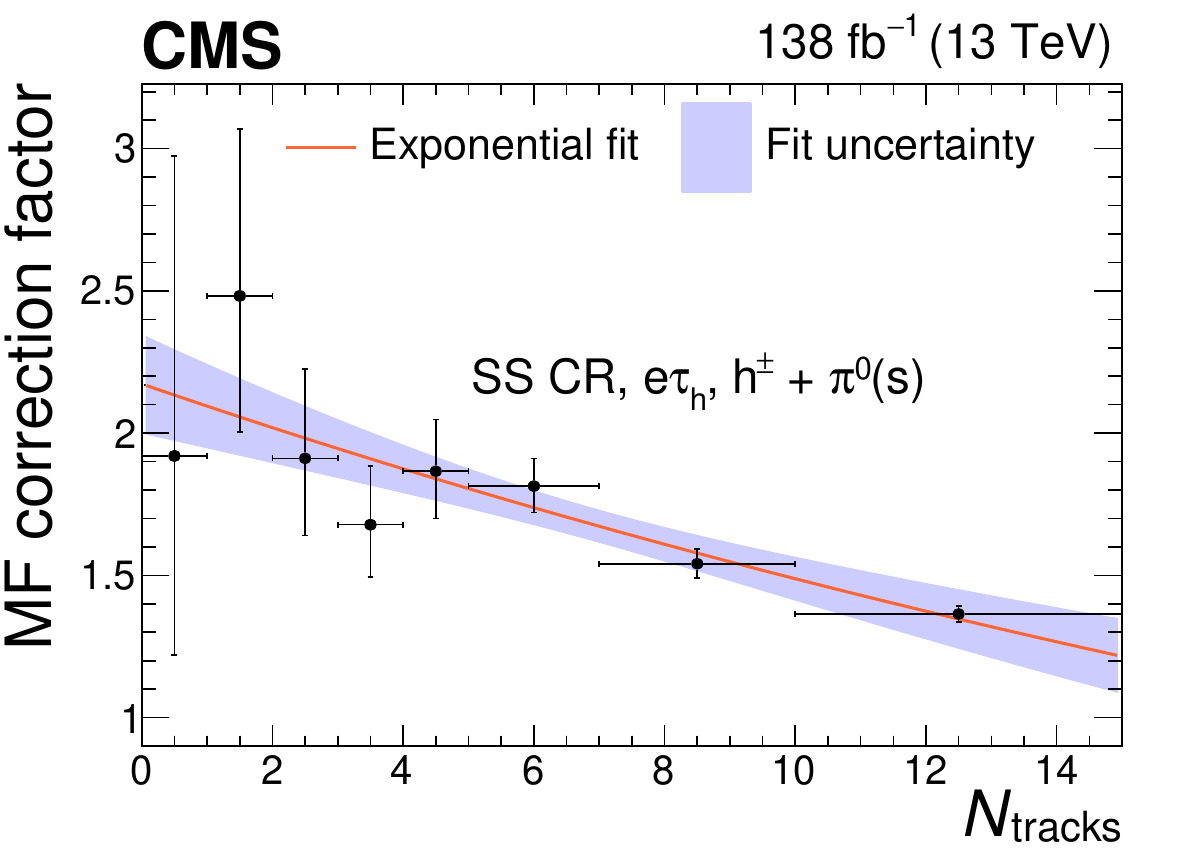}
	\caption{Multiplicative $\ntracks$-dependent corrections to the $\tauh$ MFs, $\omega(\ntracks, \DMtauh)$, in the $\etau$ final state, in the high-$\mT$ (\cmsLeft) and SS (\cmsRight) CRs, for the $\mathrm{h}^\pm+\PGpz$(s) DM. The purple shaded area corresponds to the fit uncertainty. The vertical error bars indicate the statistical uncertainty in the MF correction factors measured in individual $\ntracks$ ranges.}\label{fig:FFntracks}
\end{figure}

The track activity around the di-$\PGt$ system impacts the calculation of the $\tauh$ isolation, which enters the $\tauh$ identification discriminator, 
such that the MFs are expected to depend on $\ntracks$. For each $\tauh$ DM, the MFs determined for given ranges of $\ntracks$ are divided by the average MF measured 
inclusively in $\ntracks$ to obtain a multiplicative $\ntracks$-dependent correction, $\omega(\ntracks, \DMtauh)$,  to the \pt- and decay-mode-dependent MFs 
determined previously. The corrected MFs can then be written, for each CR, as:
	\ifthenelse{\boolean{cms@external}}{
\begin{multline}
	\MFcorr(\pttauh, \DMtauh, \ntracks) = \\ \text{MF}(\pttauh, \DMtauh)\omega(\ntracks, \DMtauh).
\end{multline}
}
	{\begin{equation}
	\MFcorr(\pttauh, \DMtauh, \ntracks) = \text{MF}(\pttauh, \DMtauh)\omega(\ntracks, \DMtauh). 
	\end{equation}}
The $\omega(\ntracks, \DMtauh)$ corrections are measured for the combination of the different data-taking years.
Because statistical uncertainties are large at low $\ntracks$, $\omega(\ntracks, \DMtauh)$ is fitted with an exponential 
function for $\ntracks$ less than 10 or 15, depending on the $\tauh$ DM, as shown in Fig.~\ref{fig:FFntracks} 
for the $\mathrm{h}^\pm+\PGpz$(s) DM in the $\etau$ final state. 

The total MF, $\MFtot$, is computed as a weighted average of the MFs determined in the high-$\mT$ and SS CRs, according to the expected relative fraction of 
$\Wjets$ and QCD multijet events:
\begin{equation}
  \MFtot = \xQCD\MFcorr_\text{SS} + (1-\xQCD)\MFcorr_\text{high-\mT},
\end{equation}
where $\xQCD = \NQCD/(\NQCD+\NW)$. The number of $\Wjets$ events passing the 
baseline selection, $\NW$, is determined from simulations, whereas the same quantity for QCD multijet events, $\NQCD$, 
is taken as the difference between data and simulated predictions for other processes in the SS CR. 

To estimate the mis-ID background in the SR in the $\etau$ and $\mutau$ final states, 
events passing the SR selection with the exception that the $\tauh$ 
candidate fails the $\tauh$ nominal identification but passes looser criteria, are reweighted with $\MFtot$. 
Contributions from events with 
real $\tauh$, electrons, or muons are estimated from simulation and subtracted from the anti-isolated region. 

\subsection{Backgrounds with misidentified jets in the \texorpdfstring{$\tau_h\tau_h$}{tauhtauh} final state}

In the $\tautau$ final state, the mis-ID background is mostly composed of QCD multijet events and the MFs are determined in 
a single CR where the $\tauh$ candidates have SS charge. The MFs for the leading (subleading) $\tauh$ are measured in events in which the 
subleading (leading) $\tauh$ passes or fails the tight identification criteria. They are in the range 0.05--0.38 depending on the $\tauh$ DM. 

The $\ntracks$-dependent correction is derived in a similar way as in the $\etau$ and $\mutau$ final states, and is between 1.8 and 2.3 
for events with $\ntracks=0$. 
In the $\tautau$ state, the number of events selected at low $\ntracks$ is larger because of the inversion of the isolation of both $\tauh$ candidates, and
contributions from exclusive dijet production become significant in the lowest $\ntracks$ bins, such that no fit is performed to the $\omega(\ntracks, \DMtauh)$
distributions. The values for $\ntracks=0,1$ are used without constraints events with $\ntracks>1$. 

In the SR, events in which the leading (subleading) $\tauh$ is a misidentified jet 
are estimated by scaling by the MFs events where the leading (subleading) $\tauh$ is anti-isolated and the subleading (leading) $\tauh$ is 
isolated. These contributions include events where both the leading and subleading $\tauh$ candidates are 
misidentified jets. To remove the double counting, events with two misidentified jets are estimated by scaling events where the two 
$\tauh$ candidates are anti-isolated with the product of the MFs for each object, and are subtracted. 

\subsection{Backgrounds with misidentified jets in the \texorpdfstring{$e\mu$}{emu} final state}

In the $\emu$ final state, the mis-ID background with a jet misidentified as an electron or a muon, dominantly composed of 
$\Wjets$ and QCD multijet events, is 
estimated from events where the electron and muon have SS charge. Events passing the SR selection with the exception that the 
electron and muon have SS charge are reweighted with a two-dimensional scale factor dependent on the electron and muon \pt. This 
factor is measured by taking the ratio of OS-to-SS events where the muon isolation is inverted, 
subtracting nonjet contributions estimated from simulation. It is in the range 1.4--2.1, depending on the 
lepton \pt. 

A bias is introduced by the inversion of the muon isolation, which modifies the background composition, 
and a correction to the scale factor is measured by taking the ratio of the scale factors measured in events where the electron identification 
is inverted, and the muon isolation is inverted or not. The correction ranges between 1.0 and 2.7 depending on the lepton \pt. 

A multiplicative correction for the dependency of the scale factor on $\ntracks$ is measured by fitting the ratio of the scale factors 
in exclusive ranges of $\ntracks$ over the average inclusive scale factor. At $\ntracks=0$, the correction is about $0.89\pm0.07$, where the uncertainty is 
mostly from statistical origin. 

Although 
the normalization of the jet mis-ID background is estimated with the procedure described above, its $\mvis$ distribution is taken from SS data events
with $\ntracks<10$, from which backgrounds other than the jet mis-ID component are subtracted using their simulated expectations. This method 
reduces statistical fluctuations in the $\mvis$ templates.

\subsection{Other backgrounds}

Minor contributions from diboson events to the SR, mostly $\PW\PW$ events, are estimated from simulation. The exclusive $\ggtoww$, 
$\ggtoee$, and $\ggtomumu$ events, which contribute mostly to the $\emu$, $\etau$, and $\mutau$ 
final states, respectively, are estimated from the elastic production simulation 
and rescaled with the same correction as the signal samples to include the dissociative components.

The agreement between data and predictions is checked in events passing the selection without the $\ntracks$ and acoplanarity requirements. 
The single top quark, $\ttbar$, and Higgs boson processes, which do not contribute to the SR with $\ntracks<2$, are included using their 
prediction from simulation.

\section{Systematic uncertainties}\label{sec:systematics}

Systematic uncertainties are considered as nuisance parameters in the statistical procedure to extract the signal significance, the signal strength $\mu$, and 
constraints on $\at$ and $\dt$. They are treated with either Gaussian (shape uncertainties) or log-normal (normalization uncertainties) 
function priors included in the likelihood function.
The signal strength $\mu$ is defined as the ratio between the observed cross section of the $\ggtotautau$ process to the theoretical cross section
predicted by the \textsc{gamma-UPC} generator for the elastic production component corrected by the multiplicative experimental scaling to include dissociative contributions.

The integrated luminosities for the 2016, 2017, and 2018 data-taking years have 1.2--2.5\% 
individual uncertainties~\cite{CMS-LUM-17-003,CMS-PAS-LUM-17-004,CMS-PAS-LUM-18-002}, 
whereas the overall uncertainty for the 2016--2018 period is 1.6\%. 
The uncertainty in the DY (diboson production) theoretical cross section is 2 (5)\%~\cite{Gavin:2012sy}. 

The uncertainties in the electron and muon identification, 
isolation, and triggering efficiencies are up to 2\% per object. Uncertainties related to the $\tauh$ trigger efficiency depend on $\DMtauh$ 
and are up to 5\%. Several uncertainties, stemming from statistical and systematic sources in the measurement, affect the $\tauh$ identification, 
with different correlations across data-taking years and various dependencies on the 
decay mode and \pt of the $\tauh$ candidate. This uncertainty is up to 5\% in the \pt range considered in this analysis, and propagates directly to the measured signal strength. 
Electrons and muons misidentified as $\tauh$ candidates, coming mostly from the $\PZ/\PGg*\to\Pe\Pe$ and $\PZ/\PGg*\to\PGm\PGm$ processes, respectively,  
have normalization uncertainties on the order of 10\%. The uncertainty in the $\tauh$ energy scale is up to 1.2\%, whereas electrons 
(muons) misidentified as $\tauh$ have energy scale uncertainties up to 5 (1)\%. 
Multiplicative corrections to the identification and isolation scale factors to account for the low-$\ntracks$ environment are derived for $\tauh$
candidates and electrons, as well as electrons and muons misidentified as $\tauh$ candidates. Scale factors for muons do not need to be 
adjusted for the low-$\ntracks$ environment. The related uncertainties are 2.1 and 2.0\% for $\tauh$
candidates and electrons, and 22 and 15\% for electrons and muons misidentified as $\tauh$ candidates, respectively. 

\begin{figure*}
  \centering
  \includegraphics[width=0.7\textwidth]{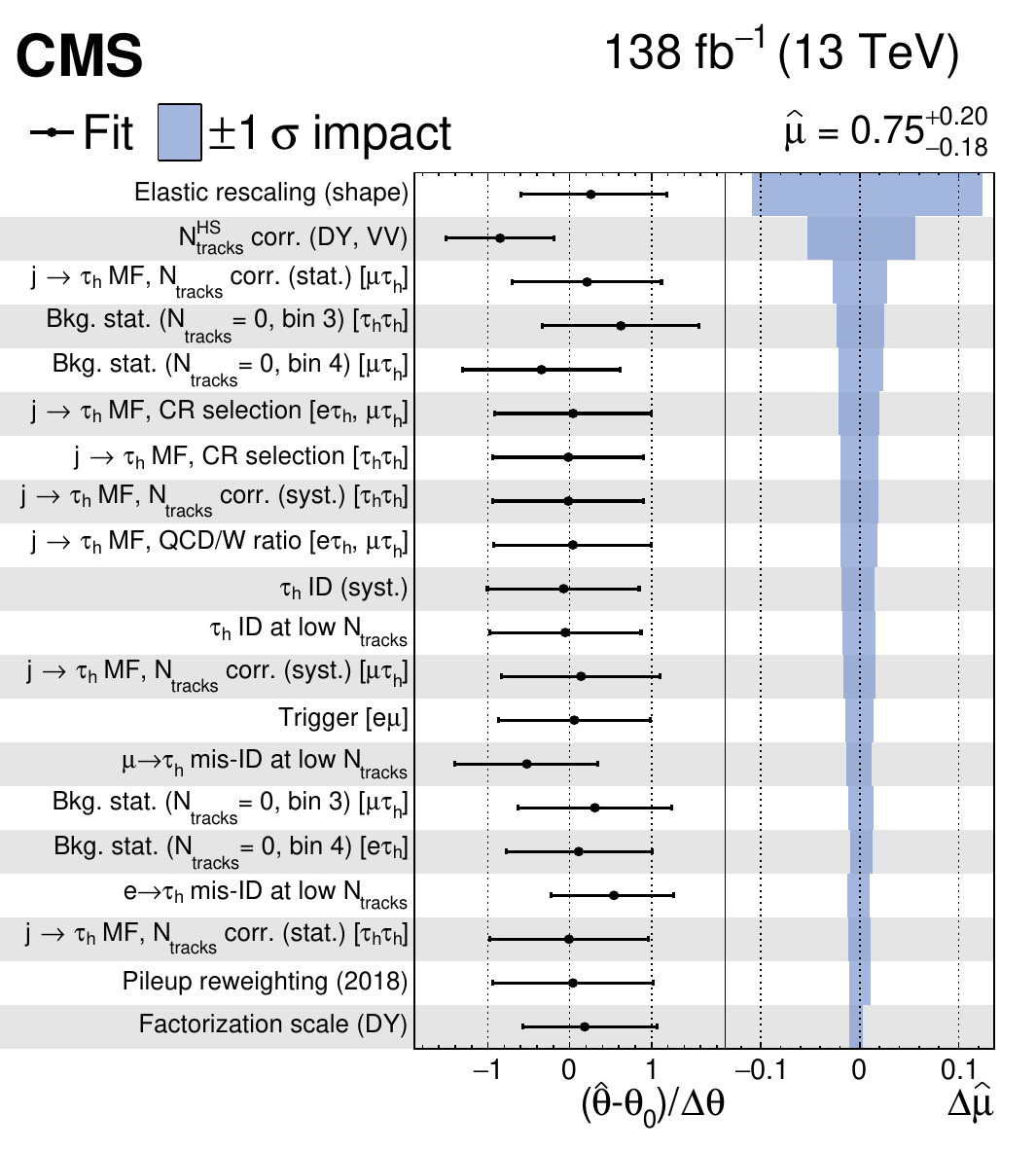}
  \caption{Postfit values of the nuisance parameters (black markers), shown as the difference of
  their best-fit values, $\hat{\theta}$, and prefit values, $\theta_0$, relative to the prefit uncertainties $\Delta\theta$. The horizontal error bars indicate the uncertainties in these measured postfit values. The impact
  $\Delta\hat{\mu}$ of the nuisance parameter on the signal strength is computed as the difference of the
nominal best fit value of $\mu$ and the best fit value obtained when fixing the nuisance parameter
  under scrutiny to its best fit value $\hat{\theta}$ plus/minus its postfit uncertainty (blue shaded area). The
nuisance parameters are ordered by their impact, and only the 20 highest ranked parameters
are shown.\label{fig:impacts} }
\end{figure*}

The uncertainty in the $\ntracksPU$ correction amounts to 2\%, which covers the effects of the beamspot width and position correction.  
The uncertainty in the $\ntracksHS$ correction is taken to have the same magnitude as the relative fraction of 
$\ggtomumu/\PW\PW$ events in the CR where the correction is derived, and 
it amounts to 6.5 (1.5)\% in the $\ntracks=0$ (1) category. The statistical uncertainty 
in the correction is negligible with respect to this systematic component. This uncertainty contributes a 7\% effect in the measured signal strength 
as shown in Fig.~\ref{fig:impacts} (second line) and is among the leading systematic uncertainties in this analysis. 

The correction to the acoplanarity distribution in 
simulated DY events leads to a 5\% normalization uncertainty, arising from the finite granularity used to bin the correction 
as a function of the \pt of the two leptons. An uncertainty of 2.0 (1.4)\% is associated with the 
estimation of the DY events in the SR with $\ntracks=0$ (1) from simulated events with $\ntracks<10$.
Uncertainties arising from the renormalization and factorization scales, as well as from the PDF set, are also taken into account for the DY simulation, 
after 
propagation to the acoplanarity correction.

The uncertainty in the predictions of the jet mis-ID background in final states with $\tauh$ candidates comes from several sources:
\begin{itemize}
  \item statistical uncertainty in the inclusive MFs, growing linearly to 50\% for $\pttauh=300\GeV$, independently for each DM and data-taking year;
  \item statistical uncertainty in the $\ntracks$ correction, amounting to 18 (16, 6)\% in the $\etau$ ($\mutau$, $\tautau$) SRs;
\item systematic uncertainty in the inclusive MFs, coming from the inversion of selection requirements to define the CRs, reaching up to 10\%;
\item systematic uncertainty in the calculation of the relative fraction of $\Wjets$ and QCD multijet events in the $\etau$ and $\mutau$ SRs, leading to a 9\% normalization effect;
\item systematic uncertainty in the $\ntracks$ correction, arising from the choice of the fit function and fit range, with an effect up to 10\% for $\ntracks=0$.
\end{itemize}

Statistical uncertainties in the OS-to-SS scaling factor used to evaluate the mis-ID background in the $\emu$ 
final state are included, independently for each bin in the electron and muon \pt where the measurement is performed. A systematic uncertainty 
of 10\% is added to cover for a potential difference in the background composition in the SR and CR. The extrapolation of the scaling 
factor to $\ntracks=0$ has a total uncertainty of 8\%, arising from the choice of the fit function and from the limited 
number of events in the measurement.

Most theoretical and experimental uncertainties cancel when renormalizing the signal prediction using the scaling factor derived from the 
$\mumu$ CR for the $\ggtomumu$ process. The uncertainty in the scaling factor is composed of a normalization component 
of statistical origin,  
amounting to 1.3 (3.7)\% for $\ntracks=0$ (1), and a component with a shape dependence on $\mtt$. 
For the latter component, we consider as a 1 standard deviation (s.d.) 
effect the hypothesis that the scale factor does not depend 
on $\mll$, and is symmetrized with respect to the nominal hypothesis of 
a linear dependence with $\mll$. This $\mtt$-dependent uncertainty is the leading uncertainty in the measurement of the $\ggtotautau$ 
process, as shown in Fig.~\ref{fig:impacts} (first line), and contributes a 16\% uncertainty to the measured signal strength. Modifying 
the $\ntracks$ range used to extract the inclusive processes in the scale factor measurement has a much smaller effect and the corresponding 
uncertainty is neglected.
These uncertainties are also considered for the photon-fusion $\PW\PW$, $\Pe\Pe$, 
and $\PGm\PGm$ production.  

Statistical uncertainties in the number of simulated background events in the SR or observed event yields in the CRs used to derive the jet mis-ID background 
are considered in all bins of the distributions using the approach from Ref.~\cite{BarlowBeeston}.
The systematic uncertainty in the pileup modeling is included by varying by $\pm$4.6\% the total inelastic cross section used to calculate simulated pileup 
distributions~\cite{Sirunyan:2018nqx}.

The impacts of the leading systematic uncertainties in the measurement of the signal strength, 
which is described in Section~\ref{sec:observation}, 
are shown in Fig.~\ref{fig:impacts}.  All the systematic uncertainties are summarized in Table~\ref{tab:systematics}.

\begin{table*}
  \topcaption{Summary of the systematic uncertainties considered in the analysis. The sources of the uncertainties, the processes they affect, and their magnitudes are indicated. }\label{tab:systematics}
  \centering
  \begin{tabular}{lll}
  Uncertainty & Process & Magnitude\\
  \hline
  Luminosity & All simulations & 1.6\% \\
  DY cross section & DY & 2\% \\
  Inclusive diboson cross section & $\PW\PW$, $\PW\PZ$, $\PZ\PZ$ & 5\% \\
  $\Pe$ ID, iso, trigger & All simulations &  up to 2\% \\
  $\Pe$ ID low-$\ntracks$ correction & All simulations & 1\% \\
  $\Pgm$ ID, iso, trigger & All simulations & ${<}2\%$ \\
  $\tauh$ ID & All simulations & 1--5\% \\
  $\tauh$ trigger & All simulations & up to 5\% \\
  $\Pe \to \tauh$ mis-ID & $\PZ/\PGg*\to\Pe\Pe$ and $\ggtoee$ & ${<}10\%$ \\
  $\Pgm \to \tauh$ ID & $\PZ/\PGg*\to\Pgm\Pgm$ and $\ggtomumu$  & ${<}10\%$ \\
  $\tauh$ energy scale & All simulations & ${<}1.2\%$ \\
  $\Pe \to \tauh$ energy scale & $\PZ/\PGg*\to\Pe\Pe$ and $\ggtoee$ & ${<}5\%$ \\
	  $\Pgm \to \tauh$ energy scale & $\PZ/\PGg*\to\Pgm\Pgm$ and $\ggtomumu$ & ${<}1\%$ \\
	  $\tauh$ ID low-$\ntracks$ correction & All simulations & 2.1\% \\
	  $\Pe$ ID low-$\ntracks$ correction & All simulations & 2.0\% \\
	  $\Pe \to \tauh$ ID low-$\ntracks$ correction & $\PZ/\PGg*\to\Pe\Pe$ and $\ggtoee$ & 22\% \\
	  $\Pgm \to \tauh$ ID low-$\ntracks$ correction & $\PZ/\PGg*\to\Pgm\Pgm$ and $\ggtomumu$ & 15\% \\
  $\ntracksPU$ reweighting & All simulations & 2\% \\
  $\ntracksHS$ reweighting & DY and inclusive VV & 1.5--6.5\% \\
  Acoplanarity correction & DY & 5\% \\
	  DY extrapolation from $\ntracks<10$ & DY simulation & 1.4--2.0\% \\
  $\mu_R$, $\mu_f$ & DY simulation & Shape \\
  PDF & DY simulation & Shape \\
	  jet$\to\tauh$ MF, extrapolation with $\pttauh$ & jet$\to \tauh$ mis-ID bkg. & ${<}50\%$ \\
	  jet$\to\tauh$ MF, $\ntracks$ extrapolation (stat.) & jet$\to \tauh$ mis-ID bkg. & 6--18\% \\
          jet$\to\tauh$ MF, inversion of CR selection & jet$\to \tauh$ mis-ID bkg. & ${<}10\%$ \\
	  jet$\to\tauh$ MF, $\xQCD$ fraction & jet$\to \tauh$ mis-ID bkg. & 9\% \\
	  jet$\to\tauh$ MF, $\ntracks$ extrapolation (syst.) & jet$\to \tauh$ mis-ID bkg. & ${<}10\%$ \\
	  jet$\to\Pe/\Pgm$ OS-to-SS (stat.) & jet$\to \Pe/\Pgm$ mis-ID bkg. & ${<}20\%$ \\
          jet$\to\Pe/\Pgm$ OS-to-SS (syst.) & jet$\to \Pe/\Pgm$ mis-ID bkg. & 10\% \\
          jet$\to\Pe/\Pgm$ OS-to-SS $\ntracks$ extrapolation & jet$\to\Pe/\Pgm$ mis-ID bkg. & 8\% \\
	  Elastic rescaling (stat.) & $\PGg \PGg \to \PGt\PGt/\Pgm\Pgm/\Pe\Pe,\PW\PW$ & 1.3--3.7\% \\
	  Elastic rescaling (syst., shape) & $\PGg \PGg \to \PGt\PGt/\Pgm\Pgm/\Pe\Pe,\PW\PW$ & Mass-dependent  \\
  Limited statistics & All processes & Bin-dependent \\
  Pileup reweighting & All simulations & Event-dependent \\
  \end{tabular}
  \end{table*}

\section{Results}

\begin{figure*}
  \centering
  \includegraphics[width=0.49\textwidth]{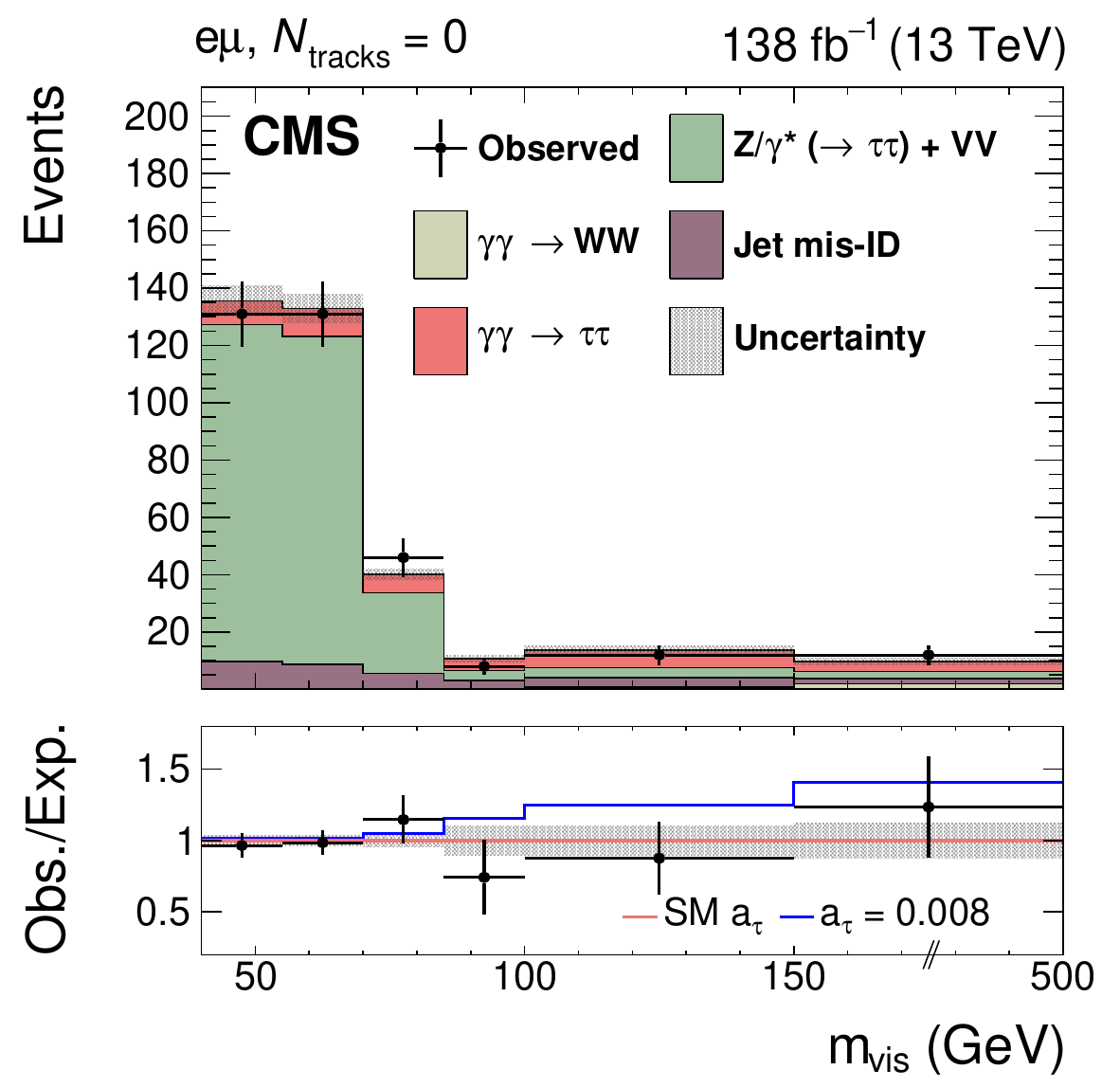}
  \includegraphics[width=0.49\textwidth]{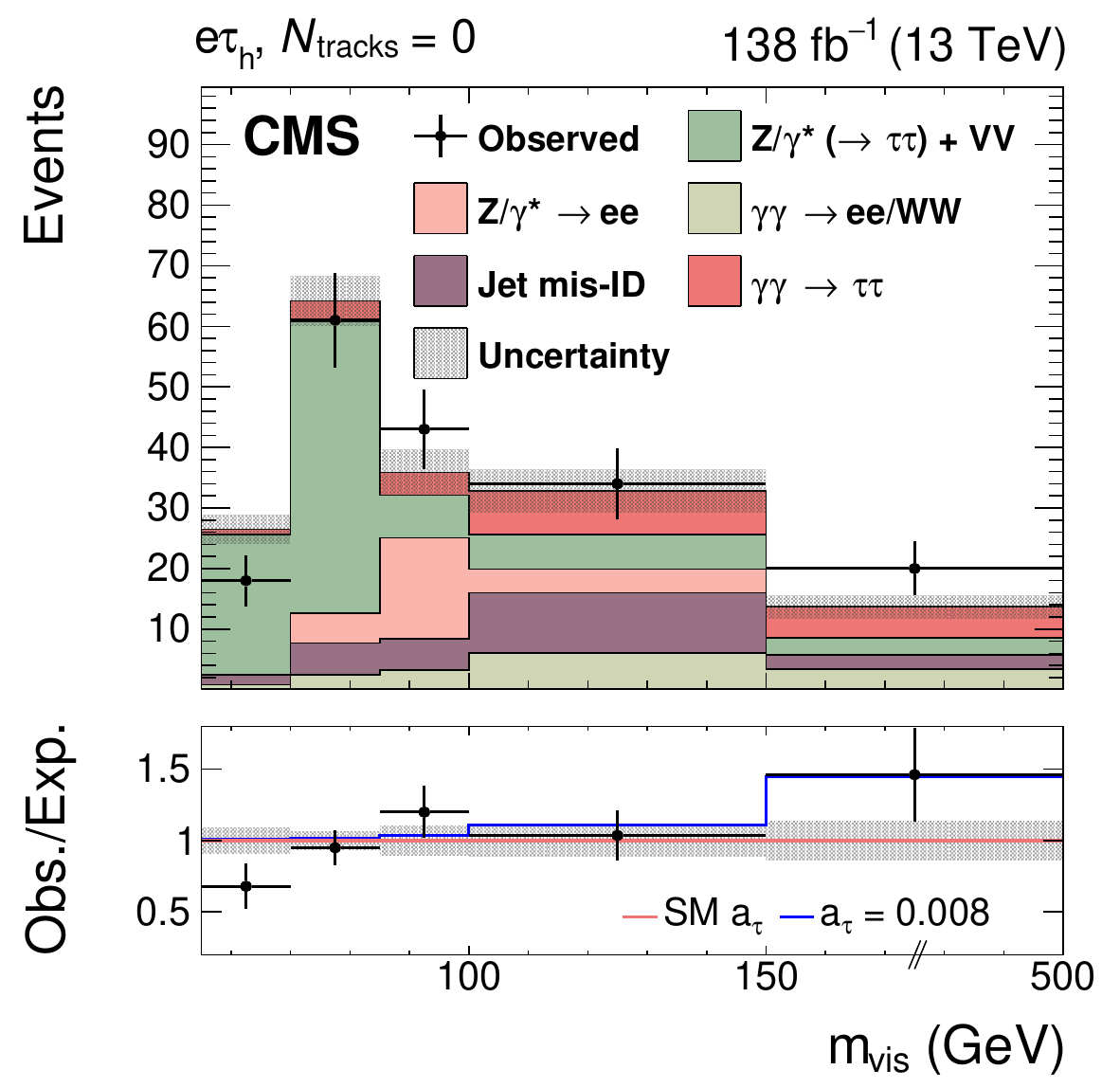}\\
  \includegraphics[width=0.49\textwidth]{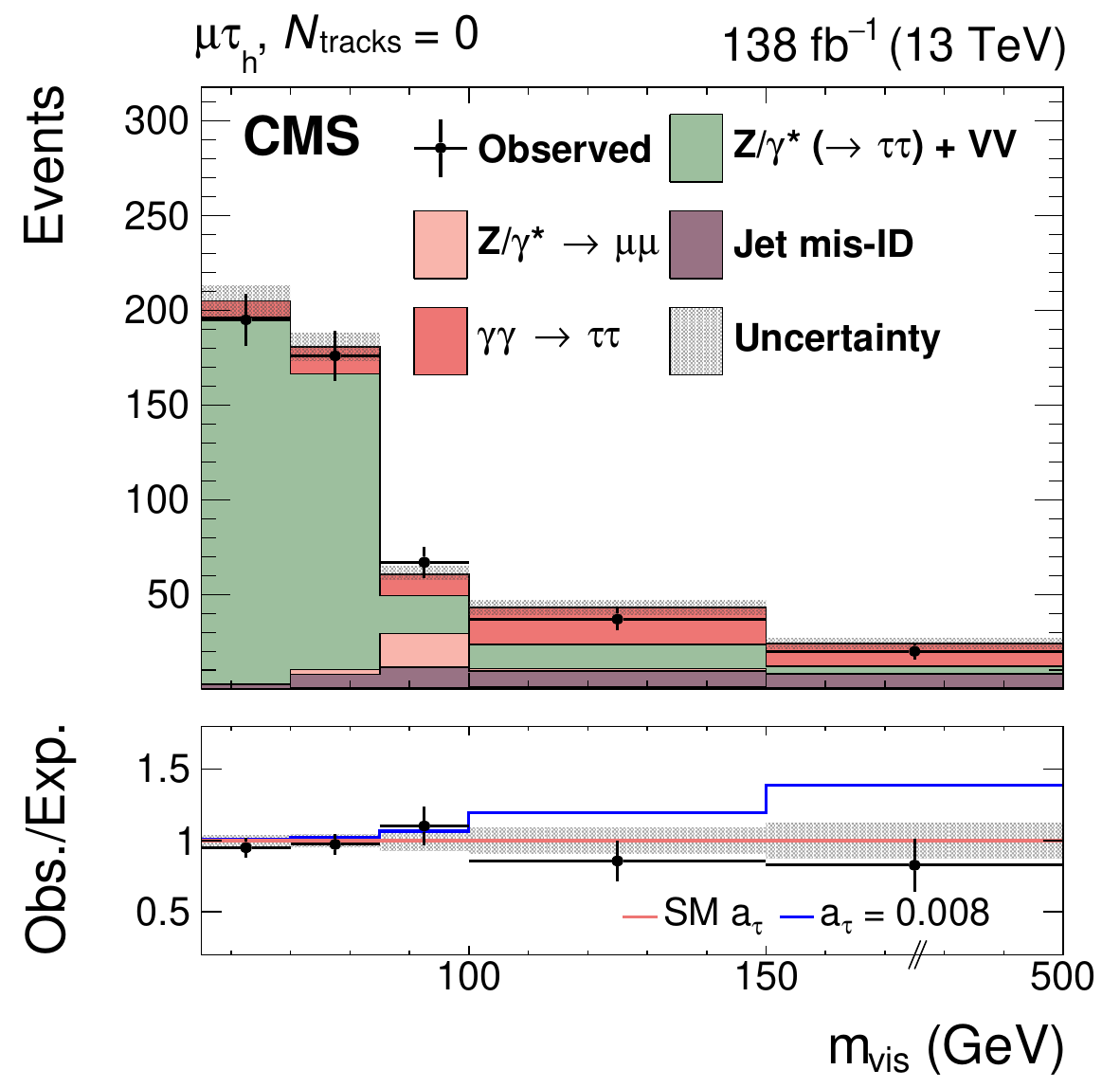}
  \includegraphics[width=0.49\textwidth]{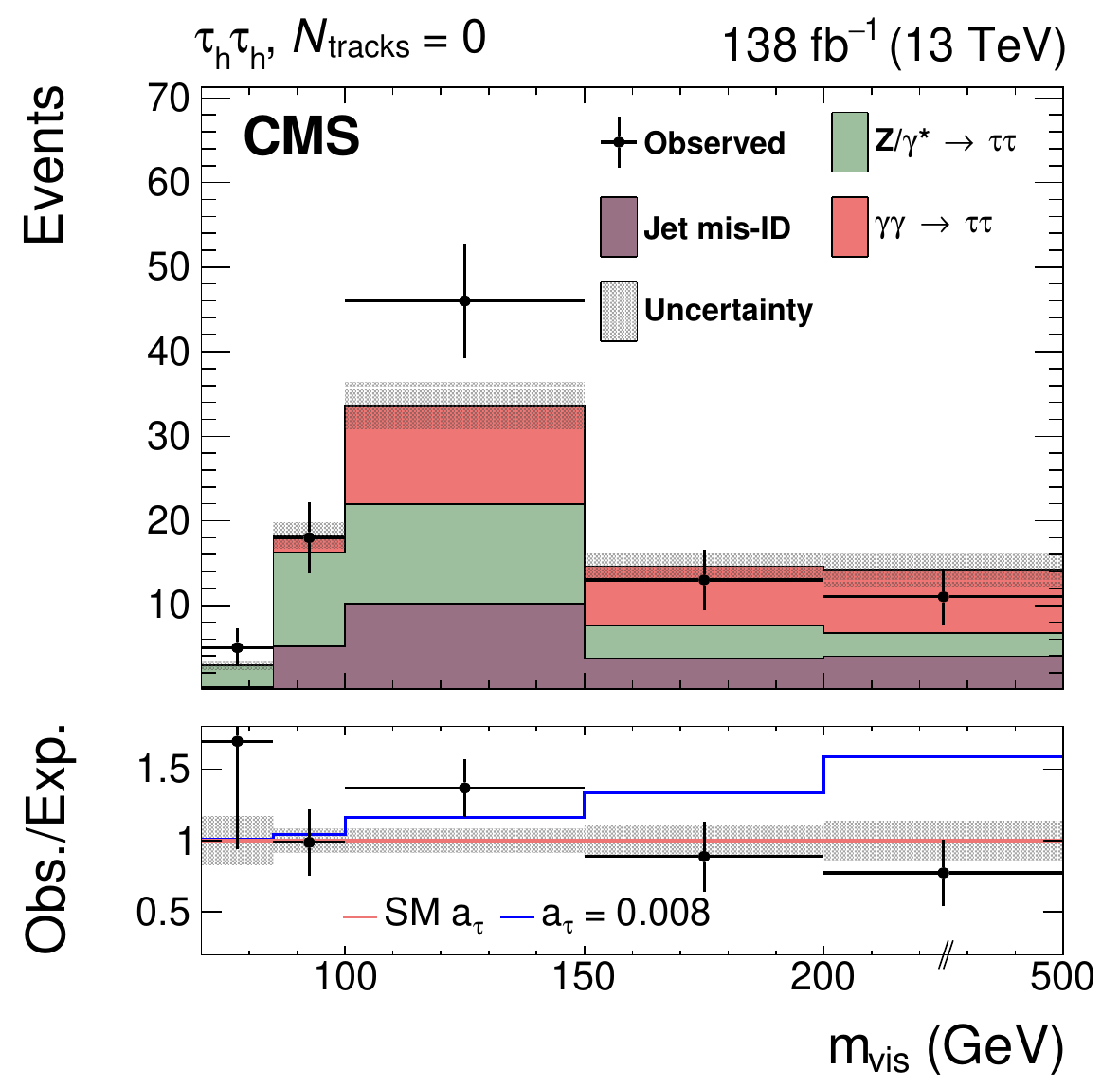}\\
  \caption{Observed and predicted $\mvis$ distributions in the $\emu$ (upper left), $\etau$ (upper right), $\mutau$ (lower left), and $\tautau$ (lower right) final states for events with $\ntracks=0$, the lower panels showing the observed/expected ratio. The observed data and their associated Poissonian statistical uncertainty are shown with black markers with vertical error bars. The minor inclusive diboson background contribution is drawn together with the DY background in the $\emu$, $\etau$, and $\mutau$ final states. The predicted background distributions correspond to the result of the global fit. The signal distribution is normalized to its best fit signal strength. The uncertainty band accounts for all sources of background and signal uncertainty, systematic as well as statistical, after the global fit. In the fit, $\at$ and $\dt$ are fixed to their SM values. The ratio of the total predictions for an illustrative value of $\at=0.008$ to those with SM electromagnetic couplings is shown with a blue line in the lower panel of each plot.\label{fig:mtt_nt0} }
\end{figure*}

\begin{figure*}
  \centering
  \includegraphics[width=0.49\textwidth]{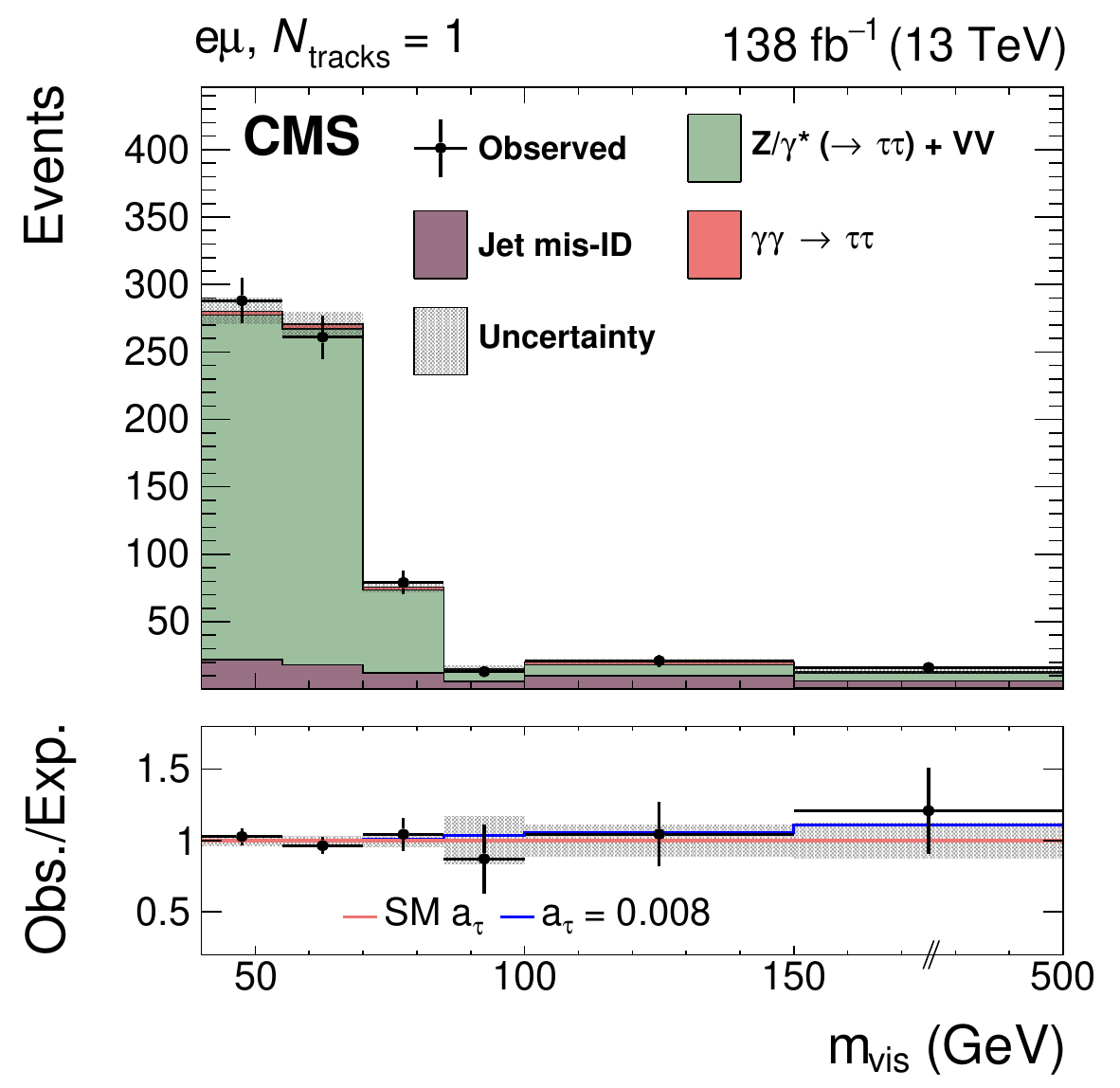}
  \includegraphics[width=0.49\textwidth]{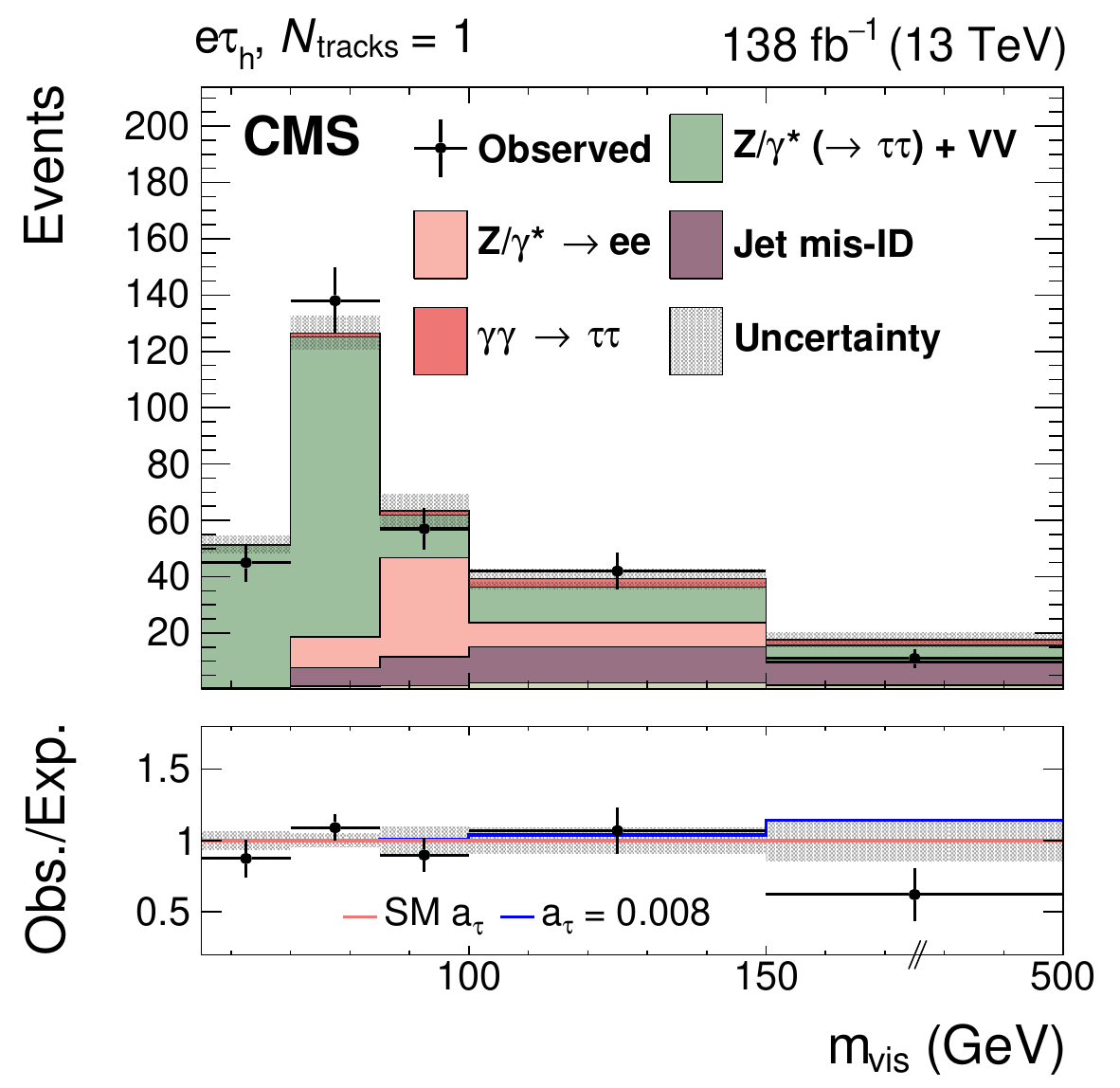}\\
  \includegraphics[width=0.49\textwidth]{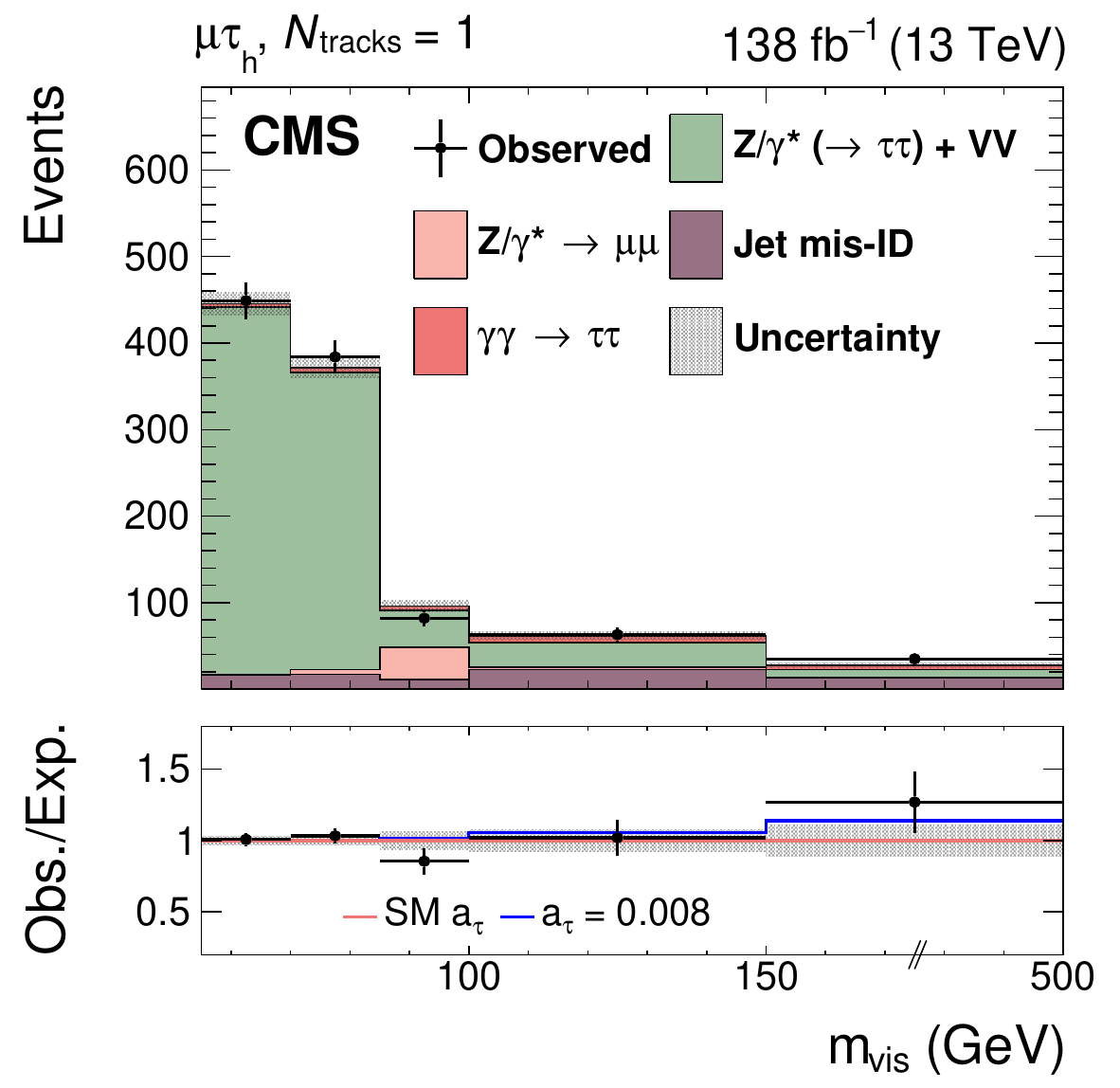}
  \includegraphics[width=0.49\textwidth]{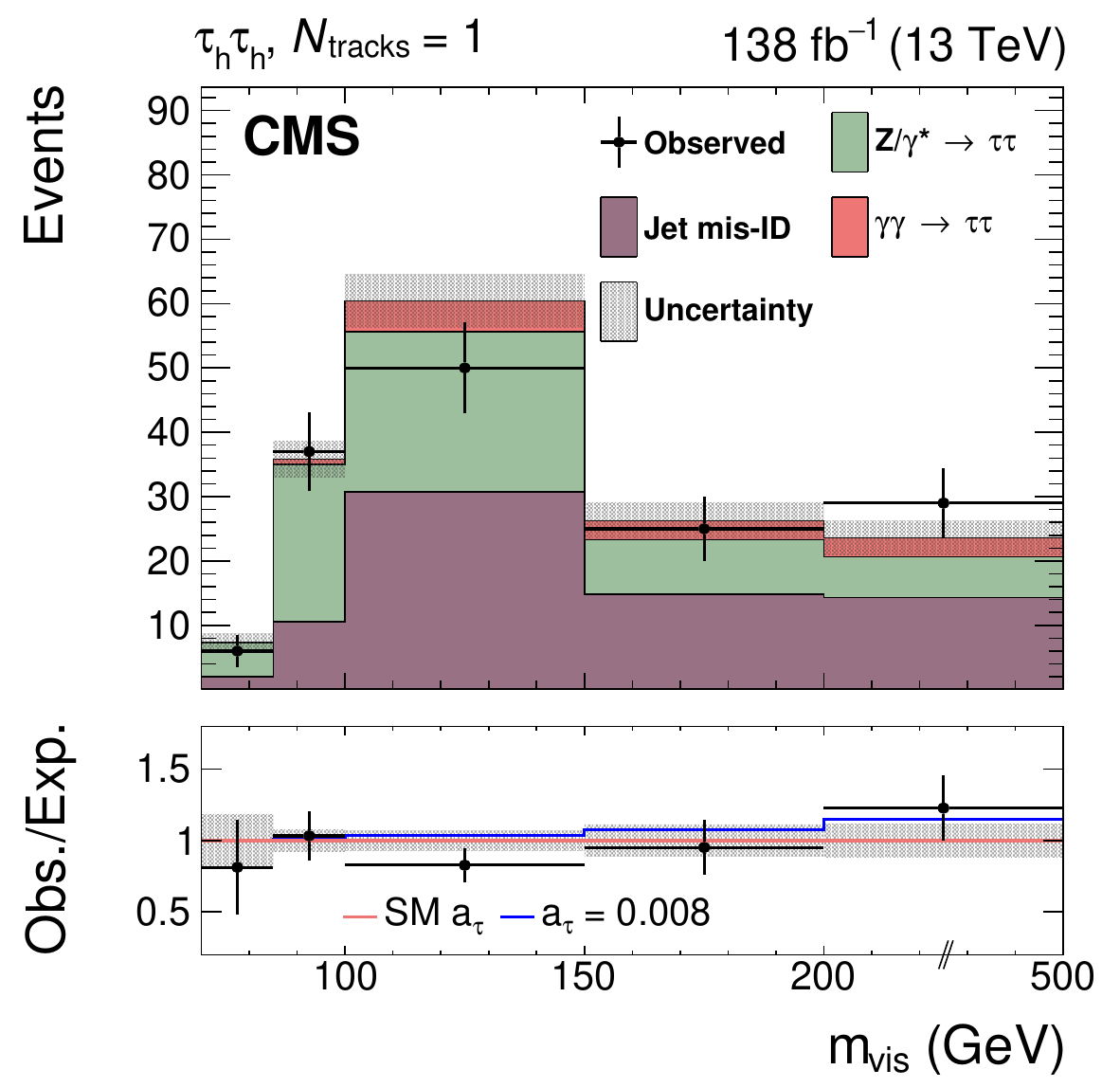}\\
  \caption{Observed and predicted $\mvis$ distributions in the $\emu$ (upper left), $\etau$ (upper right), $\mutau$ (lower left), and $\tautau$ (lower right) final states for events with $\ntracks=1$. The description of the histograms is the same as in Fig.~\ref{fig:mtt_nt0}.\label{fig:mtt_nt1} }
\end{figure*}

The significance of the $\ggtotautau$ process, as well as the constraints on $\at$ and $\dt$, are extracted with a binned maximum likelihood fit 
from the $\mvis$ distributions in the four final states 
and two $\ntracks$ bins, shown in Figs.~\ref{fig:mtt_nt0}--\ref{fig:mtt_nt1}. Since the di-$\Pgt$ system is produced centrally for the signal in the fiducial region, $\mvis$ is strongly correlated with the $\Pgt$ lepton \pt, and similarly sensitive constraints on $\at$ and $\dt$ could be 
set using $\pt$ distributions.
The CMS statistical analysis and combination tool \textsc{COMBINE} is used~\cite{CMS:2024onh}. 
The last bins include events up to $\mvis=500\GeV$, 
and events at higher $\mvis$ are discarded to ensure a sufficient difference with the $\Lambda$ scale used in the BSM interpretation. 
This upper threshold removes SM signal events at the percent level. 
The lower $\mvis$ thresholds depend on the final state because of different \pt thresholds arising from trigger requirements. 
The sources of systematic uncertainties described in Section~\ref{sec:systematics} are included in the statistical procedure as nuisance parameters. 
The number of data events in the $\ntracks=0$ SRs, with the additional requirement that $\mvis$ is greater than 100\GeV to reduce the
DY background contribution, is compared in Table~\ref{tab:yields} to the
background and signal yields in the different final states.

\begin{table*}
  \topcaption{Observed and predicted event yields per final state in the signal-enriched phase space with $\mvis>100\GeV$ and $\ntracks=0$. The signal and background yields are the result of the global fit including all sources of uncertainties. }\label{tab:yields}
\centering
\begin{tabular}{lcccc}
Process & $\emu$ & $\etau$ & $\mutau$ & $\tautau$   \\
\hline
 $\PZ/\PGg^*\to\PGt\PGt$ & $3.2 \pm  0.5$ & $8.2 \pm  1.2$ & $16.6 \pm  2.9$ & $18.4 \pm  3.1$ \\
 $\PZ/\PGg^*\to\Pe\Pe/\Pgm\Pgm$ & \NA & $4.0 \pm  1.2$ & $1.2 \pm  0.6$ & \NA \\
 Jet mis-ID & $5.2 \pm  0.8$ & $12.3 \pm  2.9$ & $15.9 \pm  3.6$ & $17.9 \pm  2.8$ \\
 Inclusive VV & $2.8 \pm  0.3$ & $0.23 \pm  0.02$ & $0.33 \pm  0.3$ & \NA \\
 $\ggtoee/\Pgm\Pgm$ & \NA & $9.2 \pm  2.3$ & $1.3 \pm  0.2$ & \NA \\
 $\ggtoww$ & $2.7 \pm  0.6$ & $0.16 \pm  0.04$ & $0.4 \pm  0.1$ & \NA \\
 Total bkg. & $13.9 \pm  1.3$ & $34.1 \pm  4.8$ & $35.7 \pm  4.4$ & $36.3 \pm  4.2$ \\
	[\cmsTabSkip]
 Signal & $9.5 \pm  2.0$ & $12.4 \pm  2.6$ & $31.6 \pm  6.7$ & $26.1 \pm  5.8$ \\
	[\cmsTabSkip]
 Total & $23.4 \pm  1.7$ & $46.5 \pm  5.4$ & $67.3 \pm  6.8$ & $62.4 \pm  6.2$ \\
 Observed & 24 & 54 & 57 & 70 \\
\end{tabular}
\end{table*}

\subsection{Observation of \texorpdfstring{$\ggtotautau$}{gamma gamma to tau tau}}\label{sec:observation}

The observed (expected) significance, calculated with the asymptotic approximation~\cite{Cowan:2010js} using \textsc{COMBINE}, is 5.3 (6.5) s.d. for the exclusive $\ggtotautau$ process. 
This constitutes the first observation of this process in $\Pp\Pp$ collisions.
The corresponding observed (expected) significances per final state are 2.3, 3.0, 2.1, and 3.4 (3.2, 2.1, 3.9, and 3.9) s.d. in the $\emu$, $\etau$, $\mutau$, and $\tautau$ final states, respectively.
We measure a best fit signal strength of $\hat{\mu}=0.75^{+0.21}_{-0.18}$, where the systematic uncertainty dominates over the statistical uncertainty ($\hat{\mu}=0.75^{+0.17}_{-0.14}\syst\pm0.11\stat$).
This corresponds to a 1.2 s.d. with respect to the theoretical cross section predicted by the \textsc{gamma-UPC} generator 
for the elastic production component corrected by the multiplicative experimental scaling to include dissociative contributions.

\begin{figure*}
  \centering
  \includegraphics[width=0.65\textwidth]{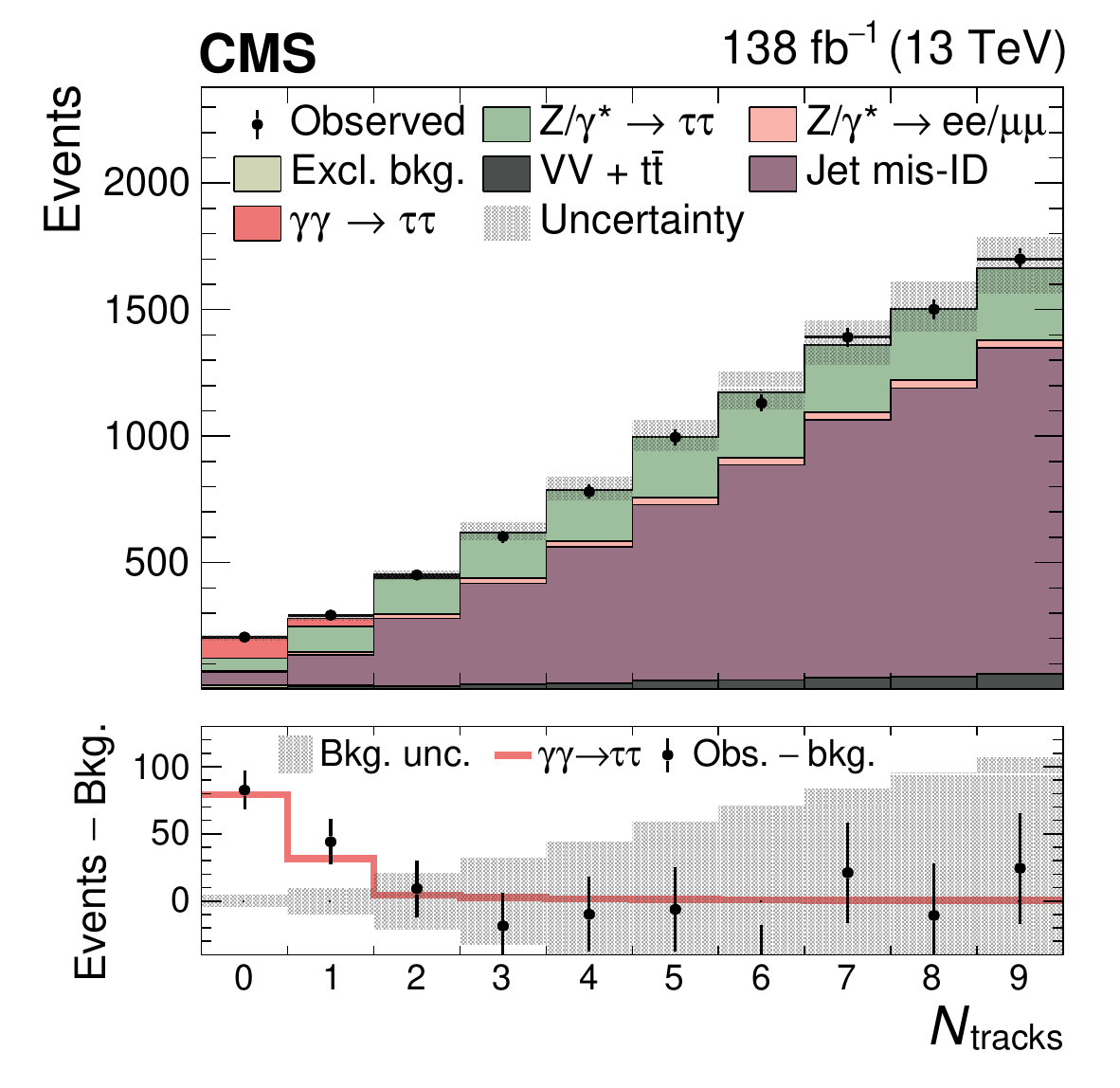}
  \caption{Observed and predicted $\ntracks$ distributions for events passing the SR selection but with the relaxed requirement $\ntracks<10$ and the additional requirement $\mvis>100\GeV$, combining the $\emu$, $\etau$, $\mutau$, and $\tautau$ final states together. The acoplanarity requirement $A<0.015$ is applied. The observed data and their associated Poissonian statistical uncertainty are shown with black markers with vertical error bars. The inclusive diboson background contribution is drawn together with that of the $\ttbar$ process. The predicted distributions are adjusted to the result of the global fit performed with the $\mvis$ distributions in the SRs, and the signal distribution is normalized to its best fit signal strength. The lower panel shows the difference between the observed events and the backgrounds, as well as the signal contribution. Systematic uncertainties are assumed to be uncorrelated between final states to draw the uncertainty band.\label{fig:nt_highmass} }
\end{figure*}

The $\ntracks$ distribution for events with $\ntracks<10$ is shown for the combination of final states in Fig.~\ref{fig:nt_highmass}
for events with $A<0.015$,
as in the SR, and $\mvis>100\GeV$, so as to reduce the DY background contribution. The signal contribution is
visible as an excess of events over the inclusive background in the first bins, whereas the agreement between prediction and observation
in the other bins demonstrates a good background modeling, more specifically in terms of  $\ntracks$ corrections to the 
simulations and MFs used to predict the background with misidentified jets. 

The fiducial cross section of the $\ggtotautau$ process is measured using only the SR with $\ntracks=0$, because it is not possible 
to extract the origin of the track in the SR with $\ntracks=1$ given the procedure used to estimate dissociative contributions. 
In this measurement, signal events not belonging to the fiducial region but entering the reconstruction-level categories are constrained 
to their expected normalizations and distributions, within uncertainties. 
The cross section of the $\ggtotautau$ process is measured in a fiducial phase space defined to be as close as possible 
to the reconstructed event
selection. All variables used in the definition of the fiducial region are calculated
at the generator level after parton showering and hadronization, and the lepton momentum
includes the momenta of photons radiated within a cone of $\Delta R < 0.1$ centered on the lepton.
The visible $\PGt$ lepton decay products at the generator level are required to have an acoplanarity $A<0.015$ and 
an invariant mass less than 500\GeV. 
The di-$\PGt$ invariant mass, including the neutrinos, is required to be greater than 50\GeV.
There 
must not be any stable charged particle with $\pt>0.5\GeV$ and $\abs{\eta}<2.5$ outside of the $\PGt$ lepton decay products. 
The other requirements depend on the final state. 
In the $\emu$ final state, the leading lepton must satisfy $\pt>24\GeV$ whereas the requirement for the 
subleading lepton is $\pt>15\GeV$. The electron (muon) must have $\abs{\eta}<2.5$ (2.4). 
In the $\etau$ ($\mutau$) final state, the electron (muon) is required to have $\pt>25\GeV$ and $\abs{\eta}<2.5$ 
($\pt>21\GeV$ and $\abs{\eta}<2.4$), whereas the visible $\tauh$ must have $\pt>30\GeV$ and $\abs{\eta}<2.3$. 
In addition, the transverse mass $\mT(\Pe/\PGm,\ptvecmiss)$ must be
less than 75\GeV.
In the $\tautau$ final state, the \pt sum of the visible $\tauh$ decay products at the generator level must be greater than 40\GeV for both $\tauh$ candidates, and the visible momentum vectors must lie within $\abs{\eta}<2.3$.
Other di-$\PGt$ final states are considered to be outside of the fiducial region. 
The definition of the fiducial region is summarized in Table~\ref{tab:fidselection}. 
Using the SRs with $\ntracks=0$, the fiducial cross section is measured to be 
$\sigma^\text{fid}_\text{obs}=12.4 ^{+3.8}_{-3.1}\unit{fb}$.
This can be compared with the prediction from the \textsc{gamma-UPC} generator for the elastic production component of the signal, 
rescaled by the correction based on control samples in the measured events to include dissociative contributions: 
$\sigma^\text{fid}_\text{pred}=16.5\pm 1.5\unit{fb}$, where the uncertainty is dominated by the data-driven  
correction.

\begin{table*}
\topcaption{Selection criteria to define the fiducial cross section. Events where the two $\PGt$ leptons decay both to electrons or to muons, with neutrinos, are considered to be outside the fiducial region. All requirements are applied using generator-level quantities, as detailed in the text. }\label{tab:fidselection}
\renewcommand{\arraystretch}{1.2}
\centering
\begin{tabular}{lllll}
 & $\emu$ & $\etau$ & $\mutau$ & $\tautau$ \\
\hline
	$\pt^{\Pe}$ (\GeVns) & $>15/24$  & $>25$ & \NA & \NA  \\
$\abs{\eta^{\Pe}}$ & $<2.5$ & $<2.5$ & \NA & \NA  \\
	$\pt^{\PGm}$ (\GeVns) & $>24/15$ & \NA & $>21$ & \NA  \\
$\abs{\eta^{\PGm}}$ & $<2.4$ & \NA & $<2.4$ & \NA  \\
	$\pttauh$ (\GeVns) & \NA & $>30$ & $>30$ & $>40$  \\
$\abs{\eta^{\tauh}}$ & \NA & $<2.3$ & $<2.3$ & $<2.3$  \\
$\Delta R(\Pell,\Pell')$ & $>0.5$ & $>0.5$ & $>0.5$ & $>0.5$  \\
	$\mT(\Pe/\PGm,\ptvecmiss)$ (\GeVns) & \NA & $<75$ & $<75$ & \NA  \\
	$A$ & $<0.015$ & $<0.015$ & $<0.015$ & $<0.015$ \\
	$\mvis$ (\GeVns) & $<500$ & $<500$ & $<500$ & $<500$ \\
	$\ntracks$ & 0 & 0 & 0 & 0 \\
\end{tabular}
\end{table*}

\subsection{Constraints on the anomalous electromagnetic moments of the \texorpdfstring{$\PGt$}{tau} lepton}

Constraints on $\at$ and $\dt$ are set independently by performing a binned likelihood scan with $\at$ and $\dt$ as single 
parameters of 
interest. Unlike in Section~\ref{sec:observation}, $\at$ and $\dt$ are allowed to float, one at a time, whereas the normalization and $\mvis$ distribution 
of the $\ggtotautau$ process under the SM hypothesis is constrained to its predicted value, estimated 
from the elastic production simulation~\cite{Shao:2022cly} and weighted with the scale factor from the $\mumu$ CR to include dissociative contributions.
Varying $\at$ and $\dt$ from their SM values modifies both the normalization of the signal process and its $\mvis$ distribution. In particular, 
the number of predicted signal events increases with $\mvis$ for large $\abs{\at}$ values,
as shown in the ratio panels of Figs.~\ref{fig:mtt_nt0}--\ref{fig:mtt_nt1} for an illustrative value of $\at=0.008$.

\begin{figure}
  \centering
  \includegraphics[width=0.49\textwidth]{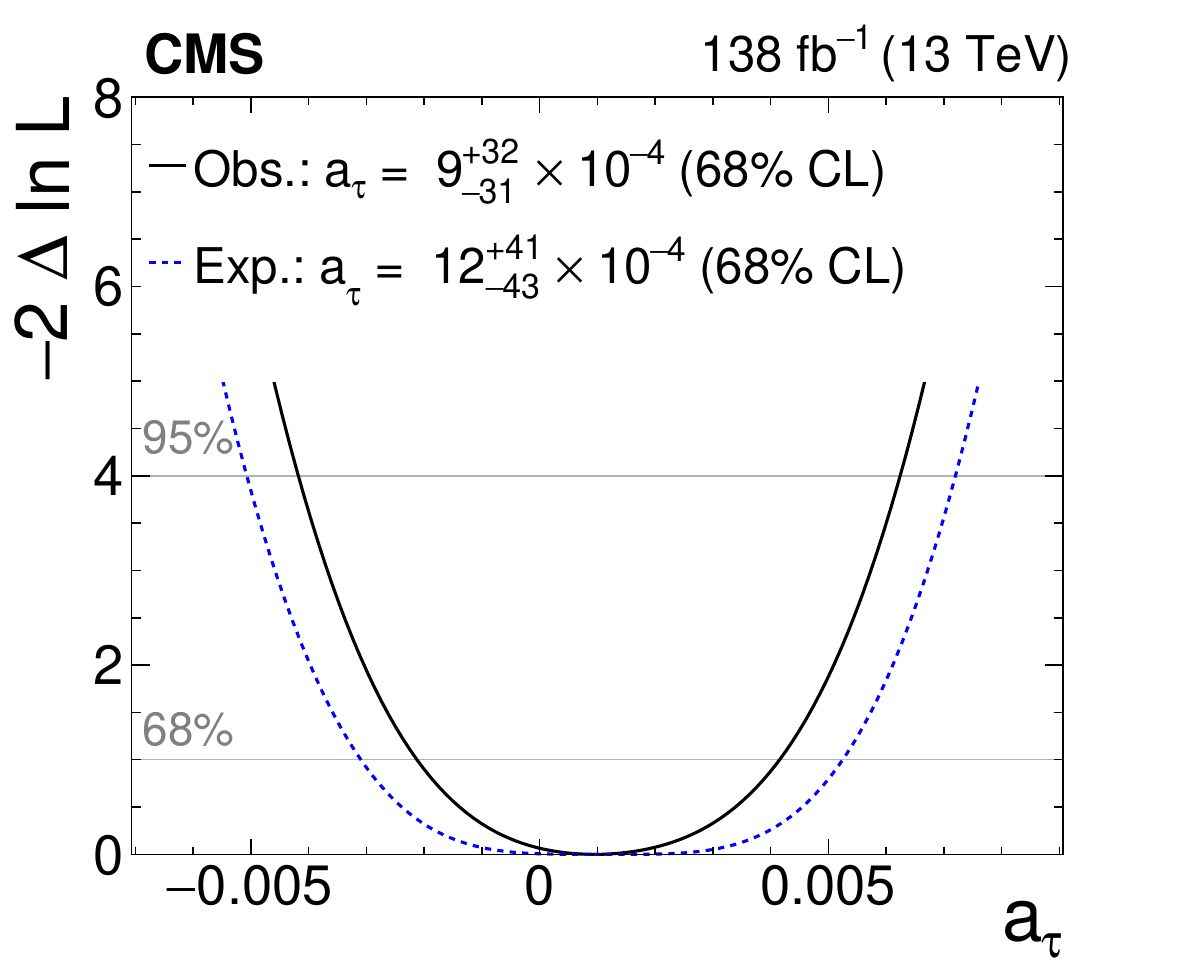}
  \includegraphics[width=0.49\textwidth]{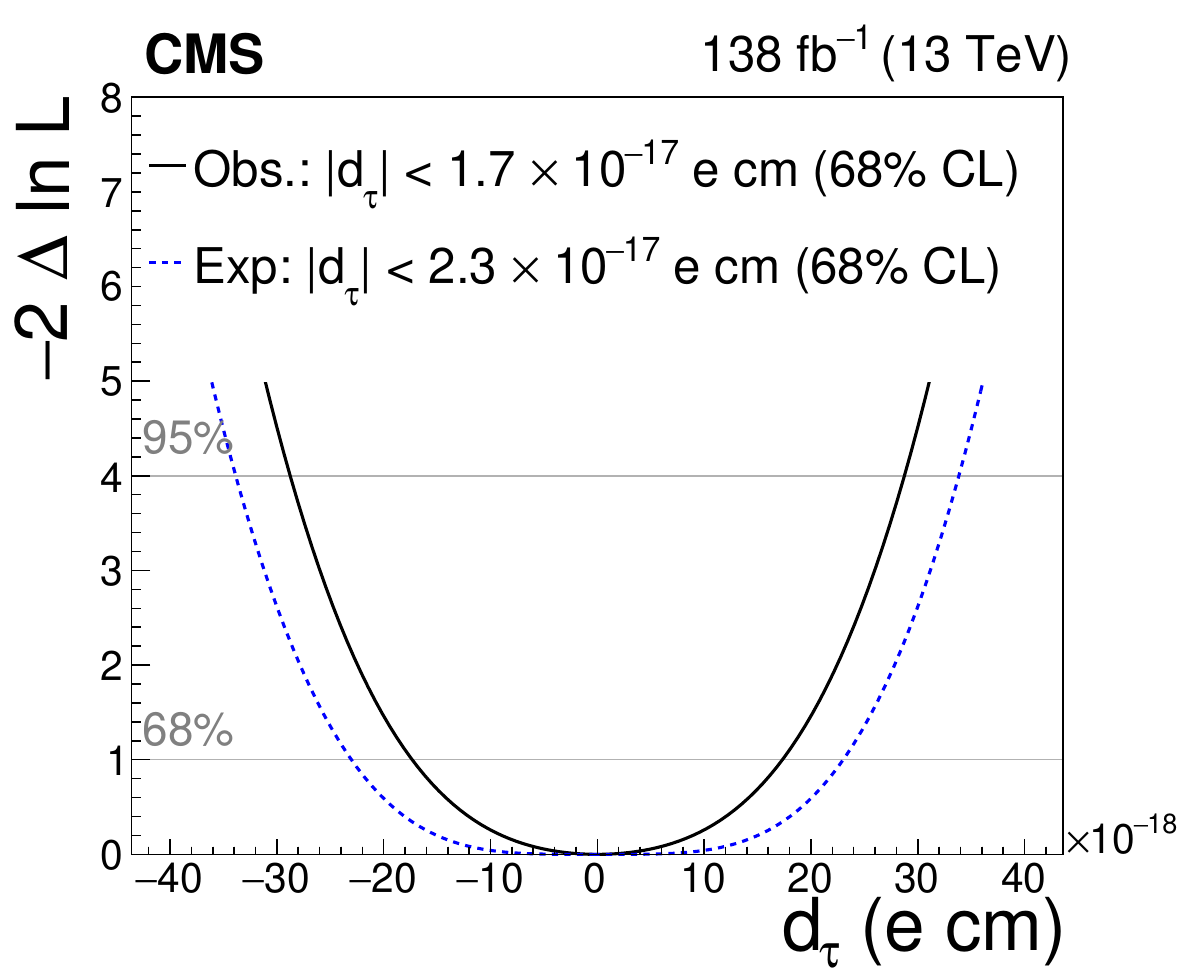}\\
  \caption{Expected and observed negative log-likelihood as a function of $\at$ (\cmsLeft) and $\dt$ (\cmsRight), for the combination of all SRs in all data-taking periods.}\label{fig:nllscan}
\end{figure}

\begin{figure}
  \centering
  \includegraphics[width=0.49\textwidth]{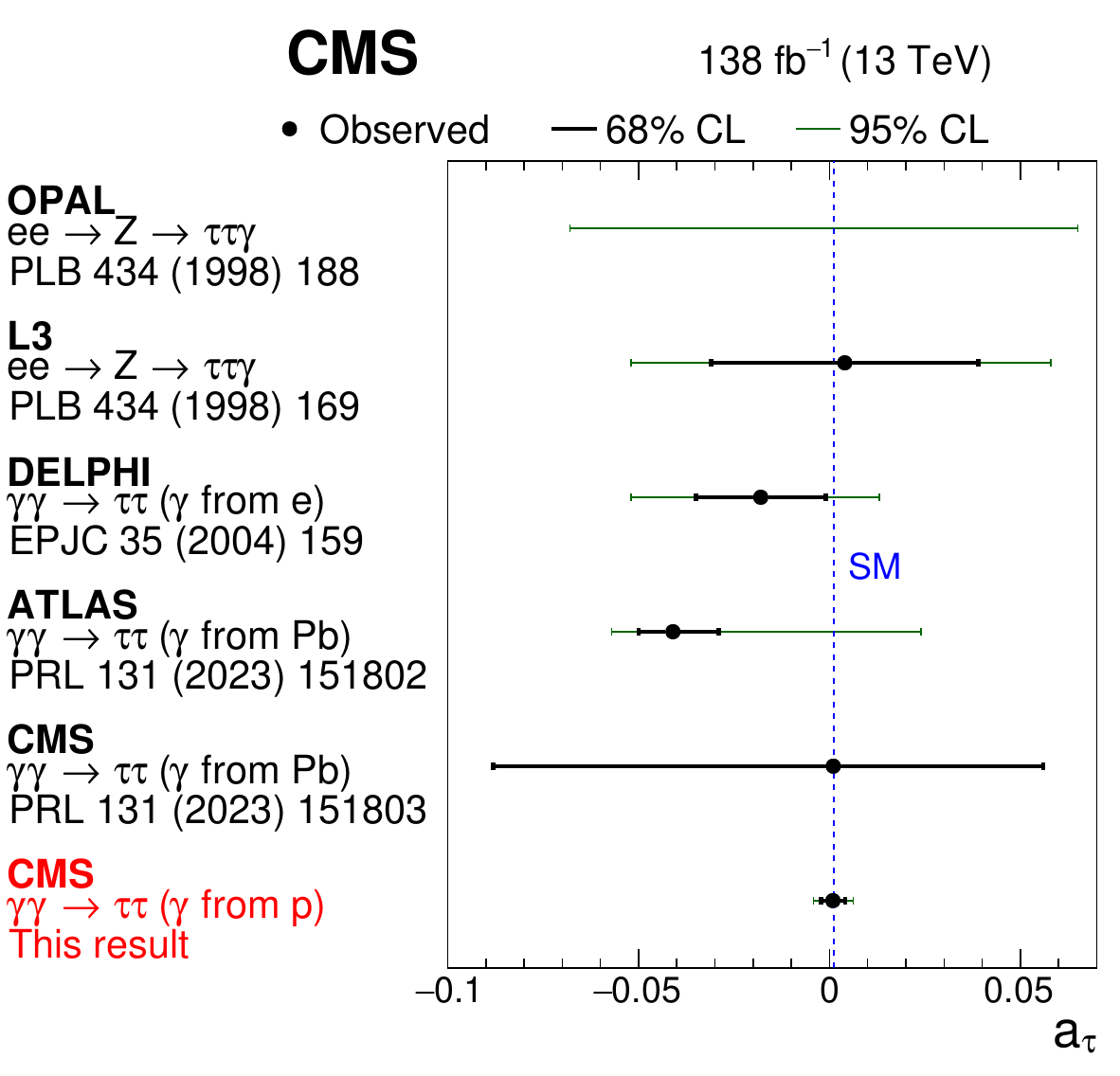}
	\includegraphics[width=0.49\textwidth]{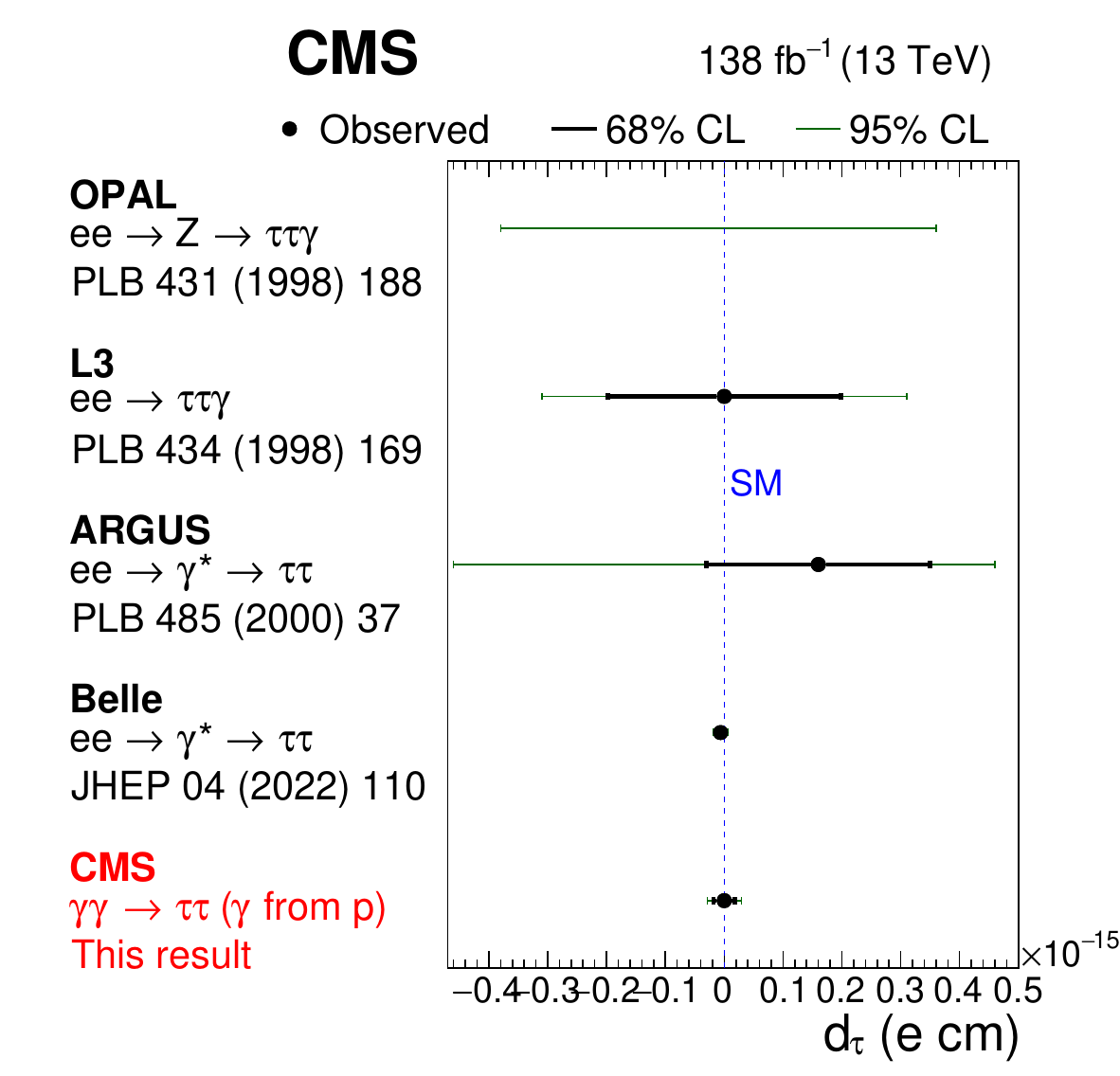}\\
	\caption{Measurements of $\at$ (\cmsLeft) and $\dt$ (\cmsRight) performed in this analysis, compared with previous results from the OPAL, L3, DELPHI, ARGUS, Belle, ATLAS, and CMS experiments~\cite{OPAL:1998dsa,L3:1998169,DELPHI:2003nah,ARGUS:2000riz,Belle:2021ybo,ATLAS:2022ryk,CMS:2022arf}. Confidence intervals at 68 and 95\% \CL are shown with thick black and thin green lines, respectively. The SM values of the $\PGt$ anomalous electromagnetic moments, $\at=1.2\times10^{-3}$ and $\dt=-7.3\times10^{-38}\ecm$, are indicated with the dashed blue lines.}
\label{fig:overview}
\end{figure}

The combination of all final states and years, using SRs with $\ntracks=0$ or 1, gives an observed best fit value of
$\at= 0.0009^{+0.0016}_{-0.0015}\syst^{+0.0028}_{-0.0027}\stat$, equivalent to $\at= 0.0009_{-0.0031}^{+0.0032}$, at 68\% confidence level (\CL). 
The corresponding expected best fit value is $\at=0.0012_{-0.0043}^{+0.0041}$, as shown in Fig.~\ref{fig:nllscan} (\cmsLeft). 
Contrary to the signal strength measurement assuming SM values for $\at$ and $\dt$, the statistical uncertainty dominates in the measurement of $\at$ 
because the sensitivity is driven by the high-$\mvis$ bins where BSM effects could be enhanced. 
The corresponding observed (expected) constraint at 95\% \CL is:
$-0.0042 < \at < 0.0062$ ($-0.0051 < \at < 0.0072$). 
The 68\% \CL constraint on $\dt$ is 
$\abs{\dt} < 1.7\times 10^{-17}\ecm$ ($\abs{\dt} < 2.3\times 10^{-17}\ecm$), with a best fit value of $\dt =0.0\times 10^{-17}\ecm$, whereas the 95\% \CL interval is 
$\abs{\dt} < 2.9\times 10^{-17}\ecm$ ($\abs{\dt} < 3.4\times 10^{-17}\ecm$).
These results are derived using events with $q^2\to 0$, which is the kinematic value at which the electromagnetic dipole moments are defined. They are compared with constraints from other experiments in Fig.~\ref{fig:overview}.
If the measurement were performed using information about the $\mvis$ distribution and with a floating normalization for the $\ggtotautau$ process, 
the expected precision in the measurement of $\at$ would decrease by about 50\%.

\begin{figure}
  \centering
  \includegraphics[width=0.49\textwidth]{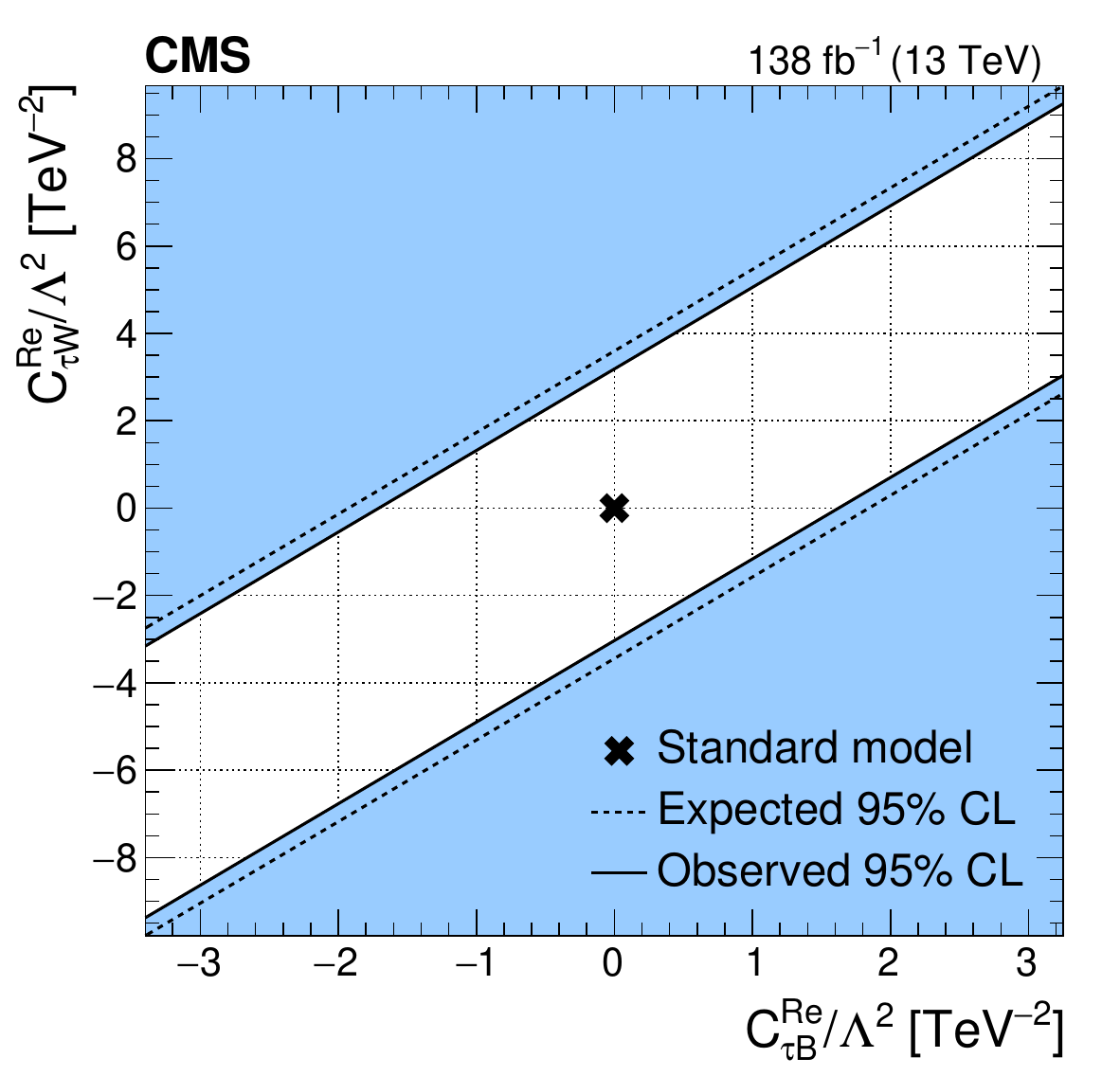}
  \includegraphics[width=0.49\textwidth]{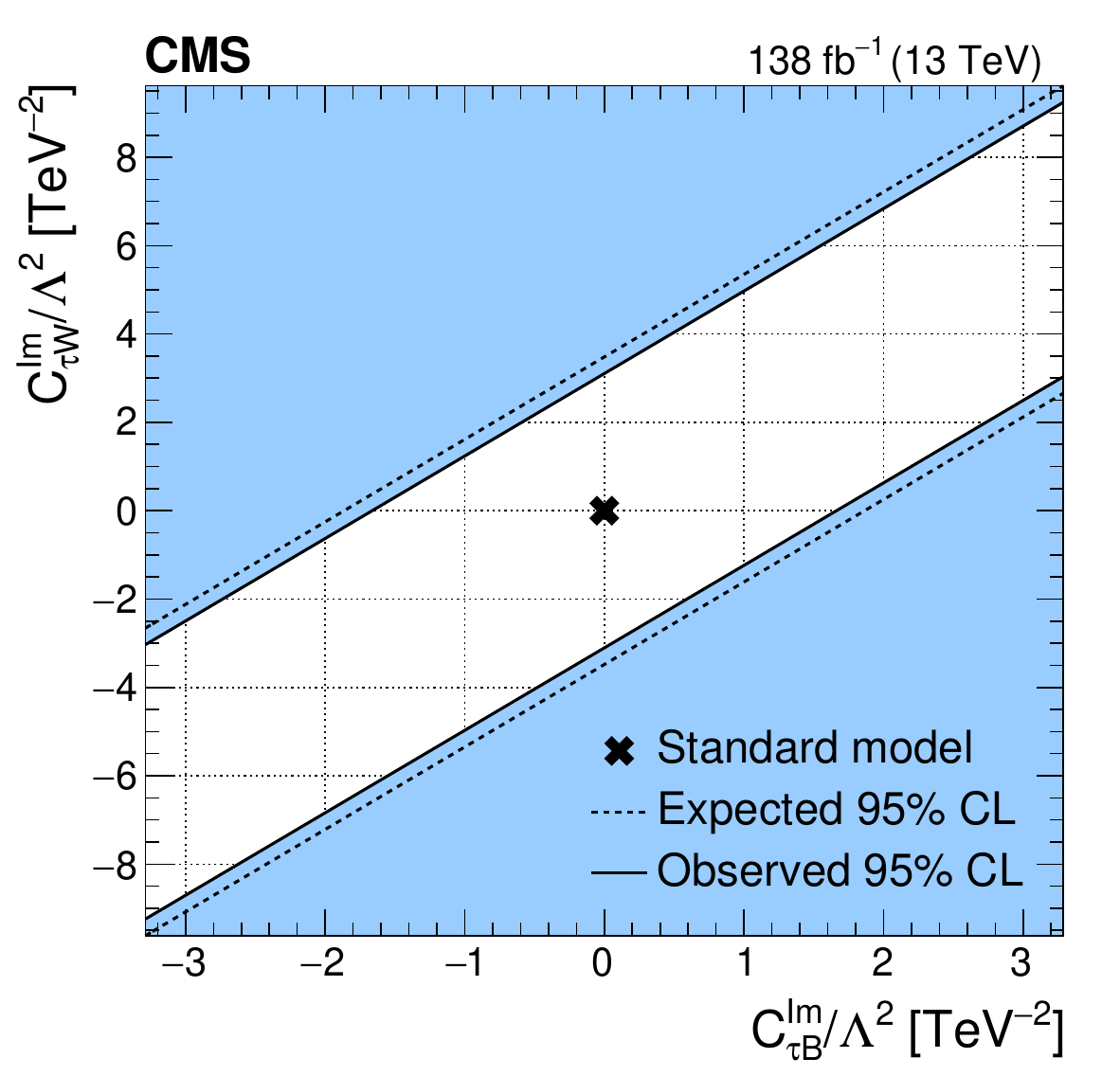}
  \caption{Expected and observed 95\% \CL constraints on the real (\cmsLeft) and imaginary (\cmsRight) parts of the Wilson coefficients $\CTB$ and $\CTW$ divided by $\Lambda^2$. The SM value is indicated with a cross. The blue shaded areas indicate excluded regions. }\label{fig:wilson}
\end{figure}

The constraints on $\at$ and $\dt$ can be converted to two-dimensional constraints on the real and imaginary parts of the Wilson coefficients $\CTB$ and $\CTW$ 
divided by $\Lambda^2$, 
using Eq.~(\ref{eq:wilson}). The 95\% \CL intervals in the plane of the normalized Wilson coefficients are shown in Fig.~\ref{fig:wilson}. 

\section{Summary}

The photon-fusion production of a pair of $\PGt$ leptons, $\ggtotautau$, has been observed for the 
first time in proton-proton collisions, with a significance of 5.3 standard deviations. The $\PGt$ leptons are reconstructed in their leptonic and hadronic decay modes. The signal has been identified by requiring low track activity around the di-$\PGt$ vertex and low azimuthal acoplanarity between 
the $\PGt$ candidates. Data in a control region with two muons were used to determine corrections for the simulations to accurately model the 
track multiplicity and to predict the signal contribution in the final state of two $\PGt$ leptons. The signal strength, fiducial cross section, and constraints on the anomalous electromagnetic moments of the $\PGt$ lepton have been extracted using the 
di-$\PGt$ invariant mass distributions in four di-$\PGt$ final states. 
The measured fiducial cross section of $\ggtotautau$ is $\sigma^\text{fid}_\text{obs}=12.4 ^{+3.8}_{-3.1}\unit{fb}$.
The anomalous $\PGt$ magnetic moment is determined to be $\at= 0.0009_{-0.0031}^{+0.0032}$, whereas the electric dipole moment of the $\PGt$ lepton 
is constrained to $\abs{\dt} < 2.9\times 10^{-17}\ecm$ at 95\% confidence level. They are both in good agreement with the predictions of the standard model of particle physics, and the measurements do not show any evidence for 
the presence of new physics that would modify the electromagnetic moments of the $\PGt$ lepton.
This is the most stringent limit on the $\PGt$ lepton magnetic moment to date, improving on the previous best constraints by nearly an order of magnitude.

\begin{acknowledgments}
We congratulate our colleagues in the CERN accelerator departments for the excellent performance of the LHC and thank the technical and administrative staffs at CERN and at other CMS institutes for their contributions to the success of the CMS effort. In addition, we gratefully acknowledge the computing centers and personnel of the Worldwide LHC Computing Grid and other centers for delivering so effectively the computing infrastructure essential to our analyses. Finally, we acknowledge the enduring support for the construction and operation of the LHC, the CMS detector, and the supporting computing infrastructure provided by the following funding agencies: SC (Armenia), BMBWF and FWF (Austria); FNRS and FWO (Belgium); CNPq, CAPES, FAPERJ, FAPERGS, and FAPESP (Brazil); MES and BNSF (Bulgaria); CERN; CAS, MoST, and NSFC (China); MINCIENCIAS (Colombia); MSES and CSF (Croatia); RIF (Cyprus); SENESCYT (Ecuador); ERC PRG, RVTT3 and MoER TK202 (Estonia); Academy of Finland, MEC, and HIP (Finland); CEA and CNRS/IN2P3 (France); SRNSF (Georgia); BMBF, DFG, and HGF (Germany); GSRI (Greece); NKFIH (Hungary); DAE and DST (India); IPM (Iran); SFI (Ireland); INFN (Italy); MSIP and NRF (Republic of Korea); MES (Latvia); LMTLT (Lithuania); MOE and UM (Malaysia); BUAP, CINVESTAV, CONACYT, LNS, SEP, and UASLP-FAI (Mexico); MOS (Montenegro); MBIE (New Zealand); PAEC (Pakistan); MES and NSC (Poland); FCT (Portugal); MESTD (Serbia); MCIN/AEI and PCTI (Spain); MOSTR (Sri Lanka); Swiss Funding Agencies (Switzerland); MST (Taipei); MHESI and NSTDA (Thailand); TUBITAK and TENMAK (Turkey); NASU (Ukraine); STFC (United Kingdom); DOE and NSF (USA).
\hyphenation{Rachada-pisek} Individuals have received support from the Marie-Curie program and the European Research Council and Horizon 2020 Grant, contract Nos.\ 675440, 724704, 752730, 758316, 765710, 824093, 101115353, 101002207, and COST Action CA16108 (European Union); the Leventis Foundation; the Alfred P.\ Sloan Foundation; the Alexander von Humboldt Foundation; the Science Committee, project no. 22rl-037 (Armenia); the Belgian Federal Science Policy Office; the Fonds pour la Formation \`a la Recherche dans l'Industrie et dans l'Agriculture (FRIA-Belgium); the Agentschap voor Innovatie door Wetenschap en Technologie (IWT-Belgium); the F.R.S.-FNRS and FWO (Belgium) under the ``Excellence of Science -- EOS" -- be.h project n.\ 30820817; the Beijing Municipal Science \& Technology Commission, No. Z191100007219010 and Fundamental Research Funds for the Central Universities (China); the Ministry of Education, Youth and Sports (MEYS) of the Czech Republic; the Shota Rustaveli National Science Foundation, grant FR-22-985 (Georgia); the Deutsche Forschungsgemeinschaft (DFG), under Germany's Excellence Strategy -- EXC 2121 ``Quantum Universe" -- 390833306, and under project number 400140256 - GRK2497; the Hellenic Foundation for Research and Innovation (HFRI), Project Number 2288 (Greece); the Hungarian Academy of Sciences, the New National Excellence Program - \'UNKP, the NKFIH research grants K 131991, K 133046, K 138136, K 143460, K 143477, K 146913, K 146914, K 147048, 2020-2.2.1-ED-2021-00181, and TKP2021-NKTA-64 (Hungary); the Council of Science and Industrial Research, India; ICSC -- National Research Center for High Performance Computing, Big Data and Quantum Computing and FAIR -- Future Artificial Intelligence Research, funded by the NextGenerationEU program (Italy); the Latvian Council of Science; the Ministry of Education and Science, project no. 2022/WK/14, and the National Science Center, contracts Opus 2021/41/B/ST2/01369 and 2021/43/B/ST2/01552 (Poland); the Funda\c{c}\~ao para a Ci\^encia e a Tecnologia, grant CEECIND/01334/2018 (Portugal); the National Priorities Research Program by Qatar National Research Fund; MCIN/AEI/10.13039/501100011033, ERDF ``a way of making Europe", and the Programa Estatal de Fomento de la Investigaci{\'o}n Cient{\'i}fica y T{\'e}cnica de Excelencia Mar\'{\i}a de Maeztu, grant MDM-2017-0765 and Programa Severo Ochoa del Principado de Asturias (Spain); the Chulalongkorn Academic into Its 2nd Century Project Advancement Project, and the National Science, Research and Innovation Fund via the Program Management Unit for Human Resources \& Institutional Development, Research and Innovation, grant B37G660013 (Thailand); the Kavli Foundation; the Nvidia Corporation; the SuperMicro Corporation; the Welch Foundation, contract C-1845; and the Weston Havens Foundation (USA).
\end{acknowledgments}
\bibliography{auto_generated} 

\cleardoublepage \appendix\section{The CMS Collaboration \label{app:collab}}\begin{sloppypar}\hyphenpenalty=5000\widowpenalty=500\clubpenalty=5000
\cmsinstitute{Yerevan Physics Institute, Yerevan, Armenia}
{\tolerance=6000
A.~Hayrapetyan, A.~Tumasyan\cmsAuthorMark{1}\cmsorcid{0009-0000-0684-6742}
\par}
\cmsinstitute{Institut f\"{u}r Hochenergiephysik, Vienna, Austria}
{\tolerance=6000
W.~Adam\cmsorcid{0000-0001-9099-4341}, J.W.~Andrejkovic, T.~Bergauer\cmsorcid{0000-0002-5786-0293}, S.~Chatterjee\cmsorcid{0000-0003-2660-0349}, K.~Damanakis\cmsorcid{0000-0001-5389-2872}, M.~Dragicevic\cmsorcid{0000-0003-1967-6783}, P.S.~Hussain\cmsorcid{0000-0002-4825-5278}, M.~Jeitler\cmsAuthorMark{2}\cmsorcid{0000-0002-5141-9560}, N.~Krammer\cmsorcid{0000-0002-0548-0985}, A.~Li\cmsorcid{0000-0002-4547-116X}, D.~Liko\cmsorcid{0000-0002-3380-473X}, I.~Mikulec\cmsorcid{0000-0003-0385-2746}, J.~Schieck\cmsAuthorMark{2}\cmsorcid{0000-0002-1058-8093}, R.~Sch\"{o}fbeck\cmsorcid{0000-0002-2332-8784}, D.~Schwarz\cmsorcid{0000-0002-3821-7331}, M.~Sonawane\cmsorcid{0000-0003-0510-7010}, S.~Templ\cmsorcid{0000-0003-3137-5692}, W.~Waltenberger\cmsorcid{0000-0002-6215-7228}, C.-E.~Wulz\cmsAuthorMark{2}\cmsorcid{0000-0001-9226-5812}
% ...
\par}
\vskip\cmsinstskip
\dag:~Deceased\\
$^{1}$Also at Yerevan State University, Yerevan, Armenia\\
$^{2}$Also at TU Wien, Vienna, Austria\\
% ...
\end{sloppypar}
%%% END EDITABLE REGION %%%
% skeleton_end
\end{document}